# Anisotropic kappa distributions I: Formulation based on particle correlations

G. Livadiotis[1*], G. Nicolaou[1], F. Allegrini[1,2]
[1] Southwest Research Institute, San Antonio, TX, USA
[2] Department of Physics and Astronomy, University of Texas at San Antonio, San Antonio, TX, USA
[*] Corresponding author: glivadiotis@swri.edu

## Abstract

We develop the theoretical basis for the connection of the variety of anisotropic distributions with the statistical correlations among particles' velocity components. By examining the most common anisotropic distribution, we derive the correlation coefficient among particle energies, show how this correlation is connected to the effective dimensionality of the velocity distribution, and derive the connection between anisotropy and adiabatic polytropic index. Having established the importance of correlation among particles in the formulation of anisotropic kappa distributions, we generalize these distributions within the framework of nonextensive statistical mechanics and based on the types of homogeneous or heterogeneous correlations among the particles' velocity components. The formulation of the developed generalized distributions mediates the main two types of anisotropic kappa distributions, considering (a) equal correlations, or (b) zero correlations, among different velocity components. Finally, the developed anisotropic kappa distributions are expressed in terms of the energy and pitch angle in arbitrary reference frames.

## 1. Introduction

The temperature is a well-defined quantity in the theoretical framework of kappa distributions and nonextensive statistical mechanics (Tsallis 2009; Livadiotis & McComas 2009; 2010). It follows the same physical definitions as in classical statistical mechanics and thermodynamics: Namely, both the kinetic (Maxwell 1866) and thermodynamic (Clausius 1862, Abe 2001, Livadiotis 2018a) definitions of temperature are equivalent, determining exactly the same physical quantity. The kinetic definition is given by the mean kinetic energy per kinetic degrees of freedom (dof), here noted by $d$, that is, the mean kinetic energy corresponding to one velocity component, $\frac{1}{2}k_B T$. While this is trivial for the isotropic distributions, where the thermal energy is the same for any dof (equidistribution theorem), in the case of anisotropic distributions the temperature is derived by the mean kinetic energy averaged over all dof:

$$\tfrac{1}{2}k_B T = \frac{\langle \varepsilon \rangle}{d} = \frac{1}{d}\sum_{i=1}^{d}\left(\tfrac{1}{2}m\left\langle u_i^{\,2}\right\rangle\right) = \frac{1}{d}\sum_{i=1}^{d}\left(\tfrac{1}{2}k_B T_i\right) \text{ , or} \tag{1a}$$

$$T = \frac{1}{d}\sum_{i=1}^{d} T_i \equiv \langle T_i \rangle_d \text{ ,} \tag{1b}$$

where the kinetic energy $\varepsilon = \frac{1}{2} m \vec{u}^2$, and its mean $\langle \varepsilon \rangle = \frac{1}{2} m \langle \vec{u}^2 \rangle$, is written setting zero bulk (or mean) velocity, $\vec{u}_{\mathrm{b}} \equiv \langle \vec{u} \rangle = 0$. The fractions of thermal components define an array of anisotropy components, $\alpha_i \equiv T_i / T$, $i$=1, … , $d$, but we are interested in the typical case of 3D distribution, where one velocity component (denoted as parallel to a reference direction, e.g., that of ambient magnetic field, $u_{\parallel}$) is different than the other velocity components (forming a perpendicular manifold, $\vec{u}_{\perp}$), with a scalar anisotropy defined by $\alpha \equiv T_{\perp} / T_{\parallel}$.

The kappa index, the parameter that labels and governs the kappa distributions, together with the temperature, constitute two intensive physical quantities, that is, independent of the size of the system and characterizing its thermodynamics. Similar to temperature, the kappa index has both thermodynamic and kinetic definitions, the former is connected with the partition of entropy (Livadiotis 2018a;b) and the latter is connected with the correlation of particle energies (Abe 1999; Livadiotis & McComas 2011; Livadiotis 2015c; 2017, Ch.5.4). Therefore, the mean of the particle kinetic energies determines the temperature, while their correlation determines the kappa index.

The correlation of particle energies has been determined for the isotropic kappa distributions. The calculation of correlation involves speed moments higher than the mean kinetic energy. It is known though that statistical moments with order higher than the second moment, which corresponds to the mean energy and the definition of temperature, diverge for sufficiently small value of the kappa index; remarkably, the calculation of correlation converges for all the values of kappa index, even though it involves moments of fourth order (see: Appendix B in Livadiotis & McComas 2011). Therefore, both the kinetic definitions of temperature (mean value of particle kinetic energy) and kappa (correlation of particle kinetic energy) are well-defined for all kappa indices.

The mean and correlation of the particle kinetic energies are given as follows:

$$\langle \varepsilon \rangle = \tfrac{1}{2} d \cdot k_{\mathrm{B}} T \;\; ; \;\; \rho = \tfrac{1}{2} d \cdot \frac{1}{\kappa} \; . \tag{2a}$$

We observe that they are both proportional to half the dof, $\frac{1}{2} d$, and to the corresponding defined thermodynamic quantity, i.e., the thermal energy and (inverse) kappa index, respectively, i.e.,

(1/2) Thermal Energy ($k_{\mathrm{B}}T$) ≡ Mean kinetic Energy per dof

(1/2) Inverse Kappa ($1/\kappa$) ≡ Correlation Coefficient per dof . (2b)

The correlation between particles' kinetic energy is expected to have an inverse relationship with the kappa index, e.g., to be proportional to the inverse of kappa. Indeed, for the Maxwell-Boltzmann case where the correlation is zero, the kappa index is infinity. On the other hand, when a physical mechanism generates a kappa distribution of particle energies/velocities, it basically acts on the connection of particles through long-range interactions that add statistical correlations among particles; for instance, the coupling between plasma constituents and the embedded magnetic field occurring on various temporal and spatial scales

"binds" particles together (Livadiotis et al. 2018). This inverse or negative relationship between correlation coefficient and kappa index is shown in Figure 1.

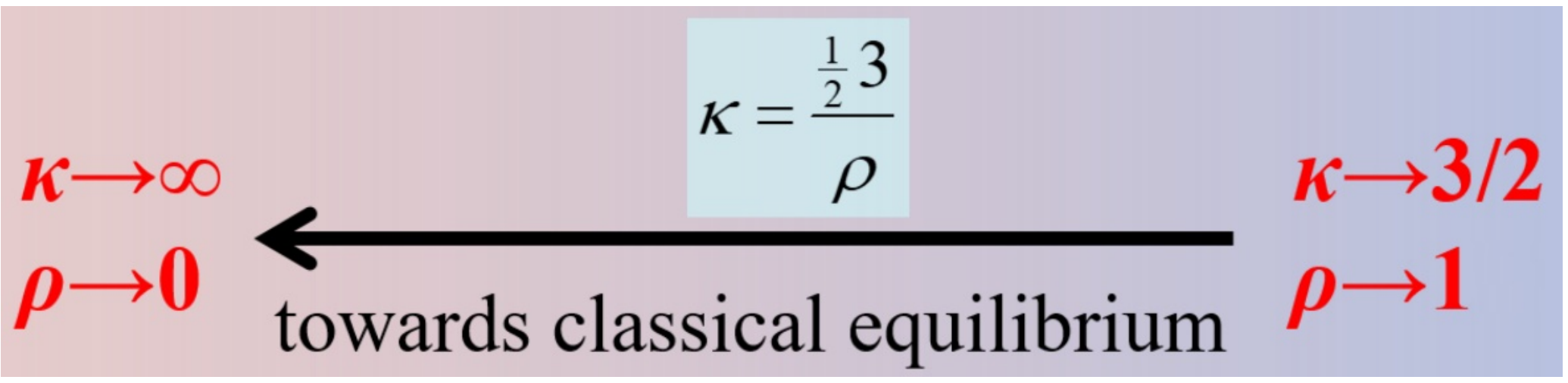

**Figure 1.** Particle systems reside in stationary states closer to the classical thermal equilibrium when the kappa index $\kappa$ increases to infinity, or equivalently, when the correlation $\rho$ decreases to zero (Livadiotis & McComas 2013).

The mean kinetic energy per dof determines the temperature, while the correlation of kinetic energy, per dof, determines the kappa index. While the mean kinetic energy per dof determines the temperature either for isotropic or anisotropic distributions, it is unknown whether the correlation per dof can consistently define the kappa index in the case of anisotropic distributions; the reason is that the correlation of kinetic energy may be expressed by a complex function of dimensionality, kappa, and anisotropy, instead of only the kappa index. In this case, the correlation per dof would have led to an ill-definition of the kappa index. Then, how would the kappa index have been defined in the case of anisotropic kappa distributions?

As we will see in this paper, the dependence on anisotropy falls into the notion of dimensionality. Let the dependence of the correlation coefficient on the dimensionality, written as $\rho=f_\rho(d)$. The involvement of the anisotropy $\alpha$ in the formulation of correlation can be understood as a deeper involvement of the anisotropy with dimensionality $d$, leading to an effective dimensionality $d_{\text{eff}}=f_d(d,\alpha)$ that substitutes the dependence of correlation, $\rho=f_\rho(d) \rightarrow \rho=f_\rho(d_{\text{eff}})=\{f_\rho \circ f_d\}(d,\alpha)$.

The impact of anisotropy on dimensionality can be shown as follows. While the overall dimensionality or dof is given by $d$=3 for any anisotropy, the limiting cases where the parallel or perpendicular directions are neglected should characterize a degeneration to 2D or 1D velocity distributions. Then, there is an effective dimensionality, $d_{\text{eff}}$, which can be defined accordingly to satisfy:

$$\begin{array}{lcl} \alpha = 0 & \Leftrightarrow & d_{\text{eff}} = 2, \\ \alpha = 1 & \Leftrightarrow & d_{\text{eff}} = 3, \\ \alpha = \infty & \Leftrightarrow & d_{\text{eff}} = 1. \end{array} \tag{3}$$

Yet, how would the correlation be formulated in the case of the anisotropic kappa distributions? How does the expression of the correlation coefficient $\rho$ for the anisotropic distributions define an effective dimensionality $d_{\text{eff}}$, which would match the dimensionalities of the limiting cases of anisotropy in Eq.(3)?

The purpose of this analysis is to (i) determine the correlation coefficient and the involved effective dimensionality of anisotropic kappa distributions characterized with homogeneous or heterogeneous correlations among their velocity components; (ii) indicate the connection of the adiabatic polytropic index with temperature anisotropy; (iii) characterize and study the types of homogeneous/heterogeneous correlations among the particles velocity components; (iv) formulate the correlation relationship that characterizes the partition of 2D joint kappa distribution into the two marginal 1D kappa distributions, as emerges from nonextensive statistical mechanics; (v) generalize the formulae of anisotropic kappa distributions, based on the various types of homogeneous/heterogeneous correlations; (vi) describe and examine the anisotropic kappa distributions in (a) the comoving reference frame with respect to the velocity components, (b) arbitrary S/C frame with respect to to the triplet of energy, pitch angle, and azimuth, and (c) the more complicate form of azimuth independent distributions with respect to energy and pitch angle.

We point out that the paper does not examine the possible mechanisms that generate the anisotropies in the velocity distribution (in the solar wind or other space plasmas, e.g., see Ao et al. 2003), but certainly it provides the impact and consequences on the particle velocity/energy distributions and the associated statistics. Note: all symbols used in this paper are defined in Table 1.

The paper is organized as follows. Section 2 briefly presents the formulation of the kappa distributions and its parameterization. In particular, we present the formulation of 3D anisotropic kappa distributions, the concept of the invariant kappa index, and the multi-dimensional anisotropic kappa distributions, which is critical for deriving the correlation coefficient. In Section 3, we calculate the correlation coefficient, that is, the normalized covariance among any two particle energies; then, we derive the effective dimensionality, which leads to the connection between anisotropy and adiabatic polytropic index. Having established the importance of correlation among particles in the formulation of anisotropic kappa distributions, in Section 4 we reverse the concept, in order to determine the variety of anisotropic formulations through their correlations; namely, we develop and study the formulae of anisotropic kappa distributions by classifying the types of correlations among the particles velocity components. These can be distinguished in homogeneous and heterogeneous correlations, where the latter type, can be further separated into a variety of different correlations that may exist among different velocity components. Section 5 shows the generalization of anisotropic kappa distributions based on the heterogeneous correlations among the particle velocity components. The generalization is developed using the partition of joint-probability distribution to its marginal distributions that helps to describe and generalize the possible types of correlations. The developed generalization of the distributions actually mediates the main two types of anisotropic kappa distributions, where the first type considers equal correlations among particles velocity components and the second type considers zero correlation among different velocity components. In Section 6, we derive the formalism of anisotropic kappa distributions in arbitrary reference frames; in particular, we transform the developed anisotropic kappa distributions to spherical coordinates with respect to the energy $E$, pitch angle $\vartheta$, and azimuth $\varphi$, as well as to the azimuth-integrated distributions with respect to energy $E$ and pitch angle $\vartheta$. Finally, Section 7 summarizes the main results of the paper. We have also written an Appendix to help understand the normalization constant of the developed distributions.

**Table 1. Involved variables and constant parameters**

| **Symbol** | **Description** |
|---|---|
| $\vec{u}$ ; $u$; $u_i$; $u_{\parallel}$, $u_{\perp}$ | Velocity vector; speed, any component; parallel and perpendicular components |
| $\vec{u}_{\mathrm{b}}$ ; $u_{\mathrm{b}}$ | Bulk velocity; bulk speed |
| $P$ | Probability distribution function |
| $\varepsilon$ | Kinetic energy |
| $E_{\mathrm{tot}\,f}$ | Kinetic energy from $N$ particles and $d$ dof per particle |
| $E$, $\vartheta$, $\varphi$ | Kinetic energy, pitch angle, azimuth, spherical coordinates in arbitrary or S/C frame |
| $E_{\mathrm{b}}$, $\vartheta_{\mathrm{b}}$, $\varphi_{\mathrm{b}}$ | Bulk kinetic energy, pitch angle, and azimuth |
| $\rho$ | Correlation coefficient (Pearson) |
| $n$ | Density |
| $T$; $T_i$; $T_{\parallel}$, $T_{\perp}$ | Temperature; any component; parallel and perpendicular components |
| $\theta$; $\theta_i$; $\theta_{\parallel}$, $\theta_{\perp}$ | Thermal speed; any component; parallel and perpendicular components |
| Α, λ | Auxiliary arguments used in the azimuth-independent distributions |
| C | Normalization constant |
| $\gamma$ | Polytropic index |
| $\alpha$ | Anisotropy |
| $\kappa$; $\kappa_d$,; $\kappa_0$ | Kappa index; kappa index for $d$-D distribution; invariant kappa index ($d$=0) |
| $\kappa_0^{\mathrm{int}}$ | Kappa index, corresponding the correlation among different velocity components |
| $d_{\mathrm{eff}}$ | Effective dimensionality |
| $d$ | Dimensionality or Degrees of freedom (dof) |
| $d_i$ | Dimensionality of the velocity vector of the $i^{\mathrm{th}}$ particle |
| $N$ | Number of correlated particles and described by a single kappa distribution |
| $f=N\cdot d$ | Total number of correlated kinetic dof |
| Γ | Gamma function |
| ${}_2F_1$, ${}_3F_2$ | Hypergeometric functions |
| $k_{\mathrm{B}}$ | Boltzmann constant |
| eV | Electron Volt |
| $m$ | Particle mass |
| $\sigma^2_{\varepsilon_1\varepsilon_2}$, $\sigma^2_{u_1^2u_2^2}$ | Covariance of the energies and velocity squares of any two particles |
| $\sigma^2_{\varepsilon\varepsilon}$, $\sigma^2_{u^2u^2}$ | Variance of the energies and velocity squares of any two particles |
| $S$ | Entropy |
| $\vec{B}$ ; $B$ | Magnetic field vector, magnitude |

## 2. Formulation of kappa distributions – a brief review

### *2.1. 3-dimensional anisotropic kappa distributions*

There is a variety of 3D anisotropic kappa distributions used for describing particle populations in space plasmas (e.g., see the primary paper of Summers & Thorne 1991, as well as, the reviews of Pierrard &

Lazar 2010; Livadiotis 2015a; see also the toolbox in the book of kappa distributions: Livadiotis 2017, Ch.4). The typical 3D anisotropic kappa distribution is given by

$$P(u_{\parallel},u_{\perp};\theta_{\parallel},\theta_{\perp},\kappa)=[\pi\,(\kappa-\tfrac{3}{2})]^{-\frac{3}{2}}\cdot\frac{\Gamma(\kappa+1)}{\Gamma(\kappa-\frac{1}{2})}\cdot\theta_{\parallel}^{-1}\theta_{\perp}^{-2}\cdot\left[1+\frac{1}{\kappa-\frac{3}{2}}\cdot\left(\frac{u_{\parallel}^{2}}{\theta_{\parallel}^{2}}+\frac{u_{\perp}^{2}}{\theta_{\perp}^{2}}\right)\right]^{-\kappa-1}\ , \tag{4a}$$

with normalization

$$\int_{0}^{\infty}\int_{-\infty}^{+\infty}P(u_{\parallel},u_{\perp};\theta_{\parallel},\theta_{\perp},\kappa)\,du_{\parallel}\,2\pi u_{\perp}du_{\perp}=1\ , \tag{4b}$$

where we have obviously considered an anisotropy with cylindrical symmetry; (for applications, see, e.g., Summers & Thorne 1991, Summers et al. 1994, Štverák et al. 2008, Astfalk et al. 2015, Khokhar et al. 2017, Wilson et al. 2019; Liu & Chen 2019; Khan et al. 2020). The involved velocity components are defined in correspondence to the direction of the magnetic field, that is, $\vec{u}_{\parallel}$ and $\vec{u}_{\perp}$ are set to be parallel (typically along the $z$-axis) and perpendicular (on the $x$-$y$ plane) to the field, respectively; the corresponding thermal speeds are explained below.

Introducing the temperature anisotropy by

$$\alpha\equiv T_{\perp}/T_{\parallel}\ , \tag{5}$$

the temperature is written as

$$T=\tfrac{1}{3}(T_{\parallel}+2T_{\perp})\ , \tag{6}$$

and the temperature-like components are expressed in terms of the actual temperature and anisotropy

$$T_{\parallel}=\frac{3}{1+2\alpha}\cdot T\ ,\ T_{\perp}=\frac{3\alpha}{1+2\alpha}\cdot T\ . \tag{7}$$

In terms of thermal speeds, these are written as

$$\theta^{2}=\tfrac{1}{3}(\theta_{\parallel}^{2}+2\theta_{\perp}^{2})\ ,\ \text{or} \tag{8}$$

$$\theta_{\parallel}^{2}=\frac{3}{1+2\alpha}\cdot\theta^{2}\ ,\ \theta_{\perp}^{2}=\frac{3\alpha}{1+2\alpha}\cdot\theta^{2}\ , \tag{9}$$

where $\theta=\sqrt{2k_{\rm B}T/m}$ denotes the thermal speed of a particle of mass $m$, that is, the temperature $T$ expressed in speed dimensions; $\theta^2/2$ provides the second statistical moment of the velocities in the co-moving reference frame. Therefore, the distribution (4a) can be written in terms of the temperature and anisotropy, i.e.,

$$P(u_{\parallel},u_{\perp};\theta,\alpha,\kappa)=\pi^{-\frac{3}{2}}(\kappa-\tfrac{3}{2})^{-\frac{3}{2}}\frac{\Gamma(\kappa+1)}{\Gamma(\kappa-\frac{1}{2})}\cdot\left(\frac{1+2\alpha}{3\alpha^{\frac{2}{3}}}\right)^{\frac{3}{2}}\theta^{-3}\cdot\left[1+\frac{1}{\kappa-\frac{3}{2}}\cdot\frac{1+2\alpha}{3\theta^{2}}\cdot\left(u_{\parallel}^{2}+\frac{1}{\alpha}u_{\perp}^{2}\right)\right]^{-\kappa-1}\ . \tag{10}$$

When the anisotropy is $\alpha$=1, then Eq.(10) recovers the standard isotropic distribution,

$$P(u;\theta,\alpha=1,\kappa)=\pi^{-\frac{3}{2}}(\kappa-\tfrac{3}{2})^{-\frac{3}{2}}\frac{\Gamma(\kappa+1)}{\Gamma(\kappa-\frac{1}{2})}\cdot\theta^{-3}\cdot\left(1+\frac{1}{\kappa-\frac{3}{2}}\cdot\frac{u^{2}}{\theta^{2}}\right)^{-\kappa-1}\ . \tag{11}$$

On the other hand, the extreme anisotropy values of $\alpha = 0$ and $\alpha \to \infty$ lead to the absolute zero of the respective temperature-like component, where the distribution in Eq.(10) is characterized by the known freezing behavior (delta function) (Livadiotis & McComas 2010), becoming narrow as a cigar (restricted along the parallel direction) or flat as a pie (restricted on the perpendicular plane), and thus, reducing the effective dimensionality (or effective degrees of freedom) to $d_{\mathrm{eff}}$=1 and $d_{\mathrm{eff}}$=2, respectively (see Eq.(3)).

Figure 2 plots the anisotropic kappa distribution, as shown in Eq.(10), for $\theta$=1, and for various values of anisotropy $\alpha$ and kappa index $\kappa$. We observe that the distribution is concentrated near $u_{\perp} = 0$ as $\alpha$ decreases tending to $\alpha = 0$ (or $T_{\perp} = 0$), and near $u_{\parallel} = 0$ as $\alpha$ increases tending to $\alpha \to \infty$ (or $T_{\parallel} = 0$). Also, similar for the isotropic case, the distribution is denser around the lines $u_{\perp} = 0$ and $u_{\parallel} = 0$ as the kappa index decreases reaching the limit of $\kappa \to \frac{3}{2}$ (or $\kappa_0 \to 0$).

## *2.2. Invariant kappa index*

The kappa index together with the temperature constitute two intensive parameters characterizing the thermodynamics of the system (Abe 2001; Livadiotis 2018a). The kappa index depends on the dimensionality, that is, the number of the correlated dof $d$, i.e., $\kappa = \kappa(d)$, also noted as $\kappa_d$. This expression is quite simple and arose from the theoretical observation that the difference $\kappa_d - \frac{1}{2}d$ is an invariant quantity independent of $d$, hence, $\kappa_d = const. + \frac{1}{2}d$. The involved constant, noted by $\kappa_0$, indicates an invariant expression of the kappa index, so that the kappa index remains invariant under variations of the dimensionality or the number of the correlated dof. Using the invariant kappa index, $\kappa_0$, the kappa distributions may be written in expressions for any dimensionality (or dof) and number of particles (Livadiotis & McComas 2011; Livadiotis 2015b).

Hereafter, we may use either the invariant kappa index $\kappa_0$, or the standard kappa index for the 3D case $\kappa_3$, accordingly, recalling that $\kappa_3 = \kappa_0 + \frac{3}{2}$. Therefore, the $d$-D kappa distribution is parametrized using either $\kappa_0$ or $\kappa_3$, by substituting $\kappa_d = \kappa_0 + \frac{1}{2}d$ or $\kappa_d = \kappa_3 + \frac{1}{2}(d-3)$, respectively. For example, the kappa index of 1D distributions is $\kappa_1$ and it can be replaced either by $\kappa_0$ or $\kappa_3$, according to $\kappa_0 = \kappa_1 - \frac{1}{2}$ or $\kappa_3 = \kappa_1 + 1$. The paper mostly focuses on the typical 3D distributions, thus we ignore the 3D subscript, writing simply $\kappa$ instead of $\kappa_3$.

The concept of invariant kappa index is necessary when dealing with (one-particle or many-particle) multi-dimensional kappa distributions.

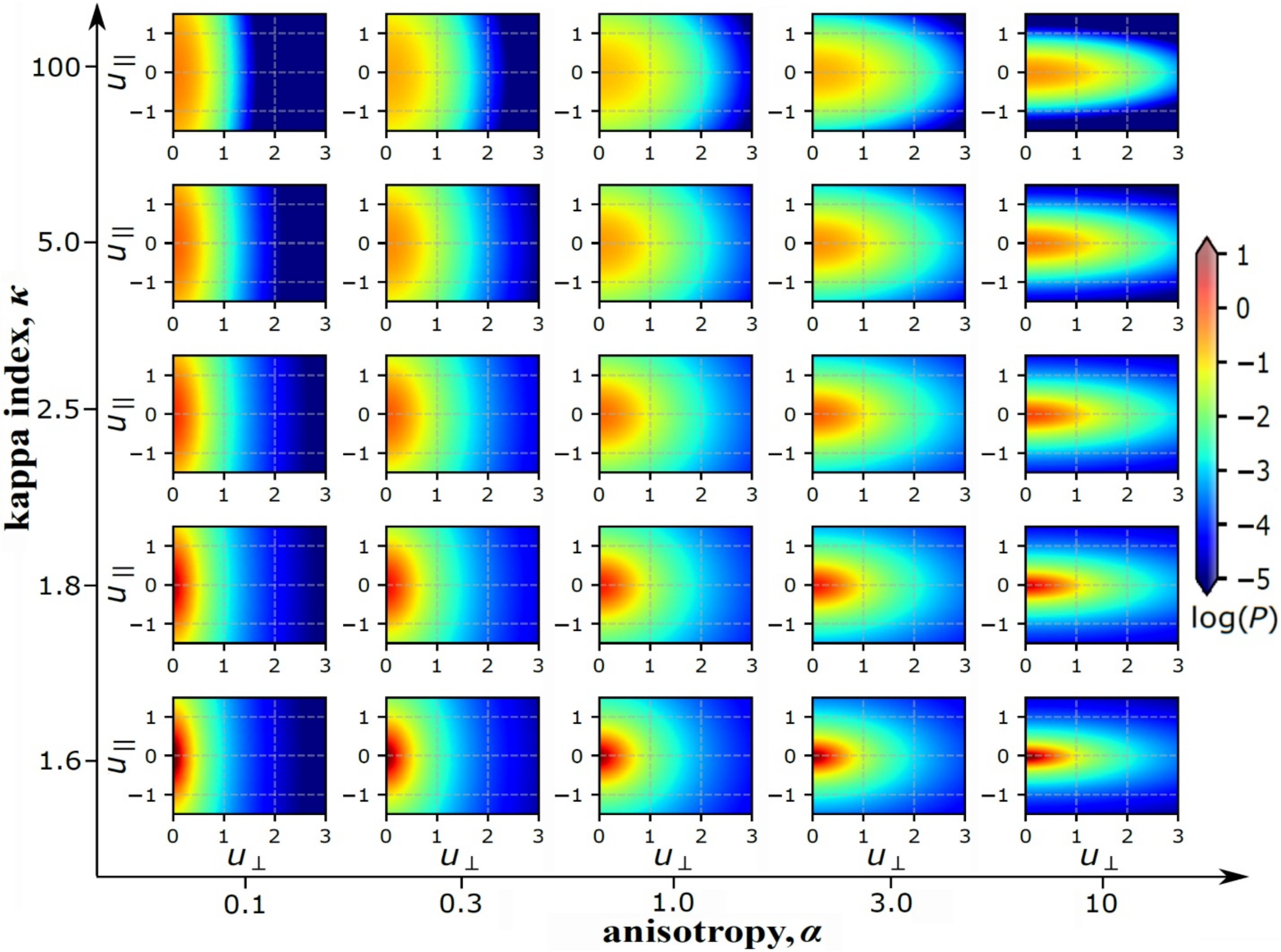

**Figure 2.** Plots of the anisotropic kappa distributions, shown in Eq.(10), for velocities normalized for $\theta$=1, and for various values of the anisotropy $\alpha$ (horizontal axis) and kappa index $\kappa$ (vertical axis). The plots with $\kappa$=100 are practically corresponding to Maxwell distributions.

*2.3. Multi-dimensional anisotropic kappa distributions*

The widely known 3D kappa distribution describes the one-particle velocities,

$$P(\vec{u};\theta,\kappa_0)=(\pi\,\kappa_0)^{-\frac{3}{2}}\cdot\frac{\Gamma(\kappa_0+\frac{5}{2})}{\Gamma(\kappa_0+1)}\cdot\theta^{-3}\cdot\left[1+\frac{1}{\kappa_0}\cdot\frac{\vec{u}^2}{\theta^2}\right]^{-\kappa_0-\frac{5}{2}}\quad, \tag{12}$$

while the *d*-D kappa distribution is written as

$$P(\vec{u};\theta,\kappa_0;d)=(\pi\,\kappa_0)^{-\frac{1}{2}d}\cdot\frac{\Gamma(\kappa_0+1+\frac{1}{2}d)}{\Gamma(\kappa_0+1)}\cdot\theta^{-d}\cdot\left[1+\frac{1}{\kappa_0}\cdot\frac{\vec{u}^2}{\theta^2}\right]^{-\kappa_0-1-\frac{1}{2}d}\quad. \tag{13}$$

In order to expose the multi-particle distribution, we have to use the dimensionality-invariance version of kappa index, $\kappa_0$. For instance the two-particle distribution is (e.g., see: Swaczyna et al. 2019),

$$P(\vec{u}_1,\vec{u}_2;\theta,\kappa_0;d)=(\pi\,\kappa_0)^{-d}\cdot\frac{\Gamma(\kappa_0+1+d)}{\Gamma(\kappa_0+1)}\cdot\theta^{-2d}\cdot\left[1+\frac{1}{\kappa_0}\cdot\frac{\vec{u}_1^{\,2}+\vec{u}_2^{\,2}}{\theta^2}\right]^{-\kappa_0-1-d}\quad. \tag{14}$$

Let the velocity vector of the *i*-th particle be $\vec{u}_i^{\,2}=u_{i,x}^{\;2}+u_{i,y}^{\;2}+u_{i,z}^{\;2}$. In a *d*-dimensional velocity space, this is $\vec{u}_i^{\,2}=u_{i,1}^{\;2}+u_{i,2}^{\;2}+\cdots+u_{i,d}^{\;2}$; therefore, the general case of a *N*-particle kappa distribution is

$$P(\vec{u}_1,\vec{u}_2,\cdots,\vec{u}_N;\theta,\kappa_0;d,N)=(\pi\,\kappa_0)^{-\frac{1}{2}Nd}\cdot\frac{\Gamma(\kappa_0+1+\frac{1}{2}Nd)}{\Gamma(\kappa_0+1)}\cdot\theta^{-Nd}\cdot\left[1+\frac{1}{\kappa_0}\cdot\frac{\vec{u}_1^{\,2}+\vec{u}_2^{\,2}+\cdots+\vec{u}_N^{\,2}}{\theta^2}\right]^{-\kappa_0-1-\frac{1}{2}Nd}\quad. \tag{15}$$

Furthermore, we follow the convention of undistinguished dof, namely, all system's correlated dof are assembled together independently of the associated particles; then, we write $\{u_i\}_{i=1}^{f}=\{u_1,\cdots,u_f\}$, $f=Nd$, meaning the notation:

$$\begin{aligned}
&u_1\equiv u_{1,1},\cdots,u_d\equiv u_{1,d}\,,\\
&u_{d+1}\equiv u_{2,1},\cdots,u_{2d}=u_{2,d}\,,\\
&u_{2d+1}=u_{3,1},\cdots,u_{3d}=u_{3,d}\,,\\
&\cdots\\
&u_{(N-1)d+1}=u_{N,1},\cdots,u_{Nd}=u_{N,d}\,.
\end{aligned} \tag{16}$$

Hence, the *N*-particle, *d*-D per particle, kappa distribution (15) can be simply expressed as an *f*-dimensional distribution (with $f=Nd$), i.e.,

$$P(u_1,u_2,\cdots,u_f;\theta,\kappa_0;f)=(\pi\,\kappa_0)^{-\frac{1}{2}f}\cdot\frac{\Gamma(\kappa_0+1+\frac{1}{2}f)}{\Gamma(\kappa_0+1)}\cdot\theta^{-f}\cdot\left[1+\frac{1}{\kappa_0}\cdot\frac{u_1^{\;2}+u_2^{\;2}+\cdots+u_f^{\;2}}{\theta^2}\right]^{-\kappa_0-1-\frac{1}{2}f}\quad. \tag{17}$$

The concept of multi-dimensional kappa distributions is necessary for determining the covariance and correlation of kinetic energies among different velocity components (see next section). Using these distributions is vital for understanding several important properties of kappa distributions, which is impossible to be done with one-particle distributions (e.g., Gravanis et al. 2020).

## 3. Derivation of correlation coefficient

The correlation coefficient of particle energy is given by the covariance of any two particle energies normalized to their energy variance, that is, respectively,

$$\sigma^2_{\varepsilon_1\varepsilon_2} = \left\langle \varepsilon_1 \varepsilon_2 \right\rangle - \left\langle \varepsilon \right\rangle^2 = \tfrac{1}{2} m \left( \left\langle u_1^{\,2} u_2^{\,2} \right\rangle - \left\langle u^2 \right\rangle^2 \right) = \tfrac{1}{2} m \sigma^2_{u_1^2 u_2^2} \ , \tag{18a}$$

and

$$\sigma^2_{\varepsilon\varepsilon} = \left\langle \varepsilon^2 \right\rangle - \left\langle \varepsilon \right\rangle^2 = \tfrac{1}{2} m \left( \left\langle u^4 \right\rangle - \left\langle u^2 \right\rangle^2 \right) = \tfrac{1}{2} m \sigma^2_{u^2 u^2} \ , \tag{18b}$$

leading to the Pearson's correlation coefficient (Abe 1999; Livadiotis & McComas 2011; Livadiotis 2015c; 2017, Ch.5.4):

$$\rho = \frac{\sigma^2_{\varepsilon_1\varepsilon_2}}{\sigma^2_{\varepsilon\varepsilon}} = \frac{\sigma^2_{u_1^2 u_2^2}}{\sigma^2_{u^2 u^2}} \ , \tag{19}$$

where all the particles have the same kinetic energy, $\left\langle \varepsilon_1 \right\rangle = \left\langle \varepsilon_2 \right\rangle = \left\langle \varepsilon \right\rangle$ .

(a) First, we derive the variance of particles' kinetic energy.

The statistical moment of the kinetic energy of order *a* is given by (e.g., Livadiotis 2017, Chapter 5.2; Livadiotis 2019a):

$$\left\langle \left( \frac{\varepsilon}{k_{\mathrm{B}} T} \right)^a \right\rangle = \kappa_0^{\ a} \cdot \frac{\Gamma(a + \frac{d}{2})}{\Gamma(\frac{d}{2})} \cdot \frac{\Gamma(\kappa_0 + 1 - a)}{\Gamma(\kappa_0 + 1)} \ , \tag{20a}$$

or, equivalently, the moment of order 2*a* of the *d*-dimensional velocity with magnitude $u = \left| \vec{u} \right|$ is

$$\left\langle \left| \vec{u} \right|^{2a} \right\rangle = \theta^a \cdot \kappa_0^{\ a} \cdot \frac{\Gamma(a + \frac{d_{\mathrm{K}}}{2})}{\Gamma(\frac{d_{\mathrm{K}}}{2})} \cdot \frac{\Gamma(\kappa_0 + 1 - a)}{\Gamma(\kappa_0 + 1)} \ . \tag{20b}$$

Then, for the parallel component (1D), we have

$$\left\langle u_{\parallel}^{\ 2} \right\rangle = \tfrac{1}{2} \theta_{\parallel}^{\ 2} \ \text{ and } \left\langle u_{\parallel}^{\ 4} \right\rangle = \tfrac{3}{4} \theta_{\parallel}^{\ 4} \cdot \frac{\kappa_0}{\kappa_0 - 1} = \tfrac{3}{4} \theta_{\parallel}^{\ 4} \cdot \frac{\kappa - \frac{3}{2}}{\kappa - \frac{5}{2}} \ , \tag{21a}$$

while for the perpendicular component (2D), we have

$$\left\langle u_{\perp}^{\ 2} \right\rangle = \theta_{\perp}^{\ 2} \ , \ \left\langle u_{\perp}^{\ 4} \right\rangle = 2 \theta_{\perp}^{\ 4} \cdot \frac{\kappa_0}{\kappa_0 - 1} = 2 \theta_{\perp}^{\ 4} \cdot \frac{\kappa - \frac{3}{2}}{\kappa - \frac{5}{2}} \ . \tag{21b}$$

We recall the expression of the covariance of two velocity components of different dimensionality (c.f., Appendix B in Livadiotis & McComas 2011),

$$\left\langle \vec{u}_i^{\ 2} \cdot \vec{u}_j^{\ 2} \right\rangle = \theta_i^2 \theta_j^{\ 2} \cdot \frac{\kappa - \frac{3}{2}}{\kappa - \frac{5}{2}} \times \begin{cases} \left( \tfrac{1}{2} d_i \right) \cdot \left( \tfrac{1}{2} d_j \right) & \text{if } i \neq j , \\ \left( \tfrac{1}{2} d_i \right) \cdot \left( \tfrac{1}{2} d_i + 1 \right) & \text{if } i = j , \end{cases} \tag{22}$$

where $d_i$ is the dimensionality of the velocity vector of the $i^{\text{th}}$ particle. Then, for the case of 1D $\vec{u}_{\parallel}$ and 2D $\vec{u}_{\perp}$ velocity components, we calculate

$$\left\langle u_{\parallel}{}^2 u_{\perp}{}^2 \right\rangle = \tfrac{1}{2}\theta_{\parallel}{}^2\theta_{\perp}{}^2 \cdot \frac{\kappa - \frac{3}{2}}{\kappa - \frac{5}{2}} \ . \tag{23}$$

Then, we derive the energy variance $\sigma^2_{u^2u^2} = \left\langle u^4 \right\rangle - \left\langle u^2 \right\rangle^2$, as follows:

$$\left\langle u^4 \right\rangle = \left\langle (u_{\parallel}{}^2 + u_{\perp}{}^2)^2 \right\rangle = \left\langle u_{\parallel}{}^4 \right\rangle + \left\langle u_{\perp}{}^4 \right\rangle + 2\left\langle u_{\parallel}{}^2 u_{\perp}{}^2 \right\rangle \ , \tag{24}$$

and substituting Eqs.(21,23), we find

$$\left\langle u^4 \right\rangle = (\tfrac{3}{4}\theta_{\parallel}^4 + 2\theta_{\perp}^4 + \theta_{\parallel}^2\theta_{\perp}^2) \cdot \frac{\kappa - \frac{3}{2}}{\kappa - \frac{5}{2}} \ . \tag{25}$$

Also, we have

$$\left\langle u^2 \right\rangle^2 = \tfrac{9}{4}\theta^4 = \tfrac{9}{4}[\tfrac{1}{3}(\theta_{\parallel}{}^2 + 2\theta_{\perp}{}^2)]^2 = \tfrac{1}{4}\theta_{\parallel}^4 + \theta_{\perp}{}^4 + \theta_{\parallel}^2\theta_{\perp}^2 \ . \tag{26}$$

Hence, from Eqs.(25,26), we find the variance

$$\sigma^2_{u^2u^2} = \left\langle u^4 \right\rangle - \left\langle u^2 \right\rangle^2 = (\tfrac{3}{4}\theta_{\parallel}^4 + 2\theta_{\perp}^4 + \theta_{\parallel}^2\theta_{\perp}^2) \cdot \frac{\kappa - \frac{3}{2}}{\kappa - \frac{5}{2}} - (\tfrac{1}{4}\theta_{\parallel}^4 + \theta_{\perp}{}^4 + \theta_{\parallel}^2\theta_{\perp}^2) \ . \tag{27}$$

(b) Next, we derive the covariance between two particle energies.

We calculate:

$$\begin{aligned} \left\langle u^2{}_1 u^2{}_2 \right\rangle &= \left\langle (u^2_{\parallel 1} + u^2_{\perp 1})(u^2_{\parallel 2} + u^2_{\perp 2}) \right\rangle \\ &= \left\langle u^2_{\parallel 1} u^2_{\parallel 2} \right\rangle + \left\langle u^2_{\perp 1} u^2_{\perp 2} \right\rangle + \left\langle u^2_{\perp 1} u^2_{\parallel 2} \right\rangle + \left\langle u^2_{\parallel 1} u^2_{\perp 2} \right\rangle \end{aligned} \tag{28}$$

where

$$\left\langle u^2_{\parallel 1} u^2_{\parallel 2} \right\rangle = \tfrac{1}{4}\theta_{\parallel}^4 \cdot \frac{\kappa - \frac{3}{2}}{\kappa - \frac{5}{2}} \ , \ \left\langle u^2_{\perp 1} u^2_{\perp 2} \right\rangle = \theta_{\perp}^4 \cdot \frac{\kappa - \frac{3}{2}}{\kappa - \frac{5}{2}} \ , \ \left\langle u^2_{\perp 1} u^2_{\parallel 2} \right\rangle = \left\langle u^2_{\parallel 1} u^2_{\perp 2} \right\rangle = \tfrac{1}{2}\theta_{\parallel}^2\theta_{\perp}^2 \cdot \frac{\kappa - \frac{3}{2}}{\kappa - \frac{5}{2}} \ , \tag{29}$$

hence,

$$\left\langle u_1{}^2 u_2{}^2 \right\rangle = (\tfrac{1}{4}\theta_{\parallel}^4 + \theta_{\perp}^4 + \theta_{\parallel}^2\theta_{\perp}^2) \cdot \frac{\kappa - \frac{3}{2}}{\kappa - \frac{5}{2}} \ . \tag{30}$$

Therefore, Eq.(28) gives the covariance:

$$\begin{aligned} \sigma^2_{u_1^2 u_2^2} &= \left\langle u_1{}^2 u_2{}^2 \right\rangle - \left\langle u^2 \right\rangle^2 \\ &= (\tfrac{1}{4}\theta_{\parallel}^4 + \theta_{\perp}^4 + \theta_{\parallel}^2\theta_{\perp}^2) \cdot \frac{\kappa - \frac{3}{2}}{\kappa - \frac{5}{2}} - (\tfrac{1}{4}\theta_{\parallel}^4 + \theta_{\perp}{}^4 + \theta_{\parallel}^2\theta_{\perp}^2) \\ &= (\tfrac{1}{4}\theta_{\parallel}^4 + \theta_{\perp}^4 + \theta_{\parallel}^2\theta_{\perp}^2) \cdot \frac{1}{\kappa - \frac{5}{2}} \ . \end{aligned} \tag{31}$$

(c) Finally, we derive the correlation.

Substituting Eqs.(27,31) in Eq.(19), we obtain

$$\rho = \frac{\sigma^2_{u_1^2 u_2^2}}{\sigma^2_{u^2 u^2}} = \frac{\frac{1}{4}\theta_{\parallel}^4 + \theta_{\perp}^4 + \theta_{\parallel}^2\theta_{\perp}^2}{(\frac{3}{4}\theta_{\parallel}^4 + 2\theta_{\perp}^4 + \theta_{\parallel}^2\theta_{\perp}^2)\cdot(\kappa - \frac{3}{2}) - (\frac{1}{4}\theta_{\parallel}^4 + \theta_{\perp}^4 + \theta_{\parallel}^2\theta_{\perp}^2)(\kappa - \frac{5}{2})}$$

$$= \frac{\frac{1}{4}\theta_{\parallel}^4 + \theta_{\perp}^4 + \theta_{\parallel}^2\theta_{\perp}^2}{(\frac{1}{2}\theta_{\parallel}^4 + \theta_{\perp}^4)\cdot(\kappa - \frac{3}{2}) + (\frac{1}{4}\theta_{\parallel}^4 + \theta_{\perp}^4 + \theta_{\parallel}^2\theta_{\perp}^2)} = \frac{\dfrac{\frac{1}{4}\theta_{\parallel}^4 + \theta_{\perp}^4 + \theta_{\parallel}^2\theta_{\perp}^2}{\frac{1}{2}\theta_{\parallel}^4 + \theta_{\perp}^4}}{(\kappa - \frac{3}{2}) + \dfrac{\frac{1}{4}\theta_{\parallel}^4 + \theta_{\perp}^4 + \theta_{\parallel}^2\theta_{\perp}^2}{\frac{1}{2}\theta_{\parallel}^4 + \theta_{\perp}^4}} \,. \tag{32}$$

Given the expressions in Eq.(9), we end up with

$$\rho = \frac{\frac{1}{2}\dfrac{(2\alpha+1)^2}{2\alpha^2+1}}{(\kappa - \frac{3}{2}) + \frac{1}{2}\dfrac{(2\alpha+1)^2}{2\alpha^2+1}}$$

$$= \frac{(\alpha + \frac{1}{2})^2}{(\alpha^2 + \frac{1}{2})(\kappa - \frac{3}{2}) + (\alpha + \frac{1}{2})^2} = \frac{(\alpha + \frac{1}{2})^2}{(\alpha^2 + \frac{1}{2})\kappa - \frac{1}{2}(\alpha - 1)^2} \,. \tag{33}$$

As stated in Eq.(2), the mean kinetic energy and correlation provide the kinetic definitions of thermal energy, $k_{\mathrm{B}}T$ , and the inverse kappa index, $1/\kappa$ , respectively. The latter is rewritten as an expression of the dimensionality and the invariant kappa index $\kappa_0 = \kappa - \frac{3}{2}$, i.e.,

$$\rho = \frac{\frac{1}{2}d}{\kappa_0 + \frac{1}{2}d} \text{ or } \rho = \frac{\frac{1}{2}d}{(\kappa - \frac{3}{2}) + \frac{1}{2}d} \,. \tag{34}$$

In the case of anisotropic distributions, the effective dimensionality interwoven with the correlations among particles should be different from the dimensionality of the embedded 3D velocity space; instead, it should follow the limiting cases stated in Eq.(3).

The effective dimensionality is derived from comparing the correlation of particle kinetic energies for a given anisotropy $\alpha$,

$$\rho = \frac{\dfrac{\frac{1}{4}\theta_{\parallel}^4 + \theta_{\perp}^4 + \theta_{\parallel}^2\theta_{\perp}^2}{\frac{1}{2}\theta_{\parallel}^4 + \theta_{\perp}^4}}{(\kappa - \frac{3}{2}) + \dfrac{\frac{1}{4}\theta_{\parallel}^4 + \theta_{\perp}^4 + \theta_{\parallel}^2\theta_{\perp}^2}{\frac{1}{2}\theta_{\parallel}^4 + \theta_{\perp}^4}} = \frac{\frac{1}{2}\dfrac{(2\alpha+1)^2}{2\alpha^2+1}}{(\kappa - \frac{3}{2}) + \frac{1}{2}\dfrac{(2\alpha+1)^2}{2\alpha^2+1}} \,, \tag{35}$$

which is written in terms of dimensionality as

$$\rho = \frac{\frac{1}{2}d_{\mathrm{eff}}}{\kappa_0 + \frac{1}{2}d_{\mathrm{eff}}} \text{ or } \rho = \frac{\frac{1}{2}d_{\mathrm{eff}}}{(\kappa - \frac{3}{2}) + \frac{1}{2}d_{\mathrm{eff}}} \,, \tag{36}$$

where the effective dimensionality (or effective dof) is determined by

$$\tfrac{1}{2}d_{\mathrm{eff}} = \frac{\frac{1}{4}\theta_{\parallel}^4 + \theta_{\perp}^4 + \theta_{\parallel}^2\theta_{\perp}^2}{\frac{1}{2}\theta_{\parallel}^4 + \theta_{\perp}^4} = \frac{1}{2}\frac{(2\alpha+1)^2}{2\alpha^2+1} \text{ , or} \tag{37}$$

$$d_{\text{eff}}(\alpha) = \frac{(2\alpha+1)^2}{2\alpha^2+1} \quad . \tag{38}$$

Next, we connect the expression in Eq.(38) with the polytropic index that describes adiabatic thermodynamic processes. The adiabatic polytropic index is given in terms of the effective kinetic degrees of freedom (Sckopke et al. 1981; Newbury et al. 1995; Kartalev et al. 2006; Nicolaou et al. 2014a; 2014b; 2015; Pang et al. 2015a; 2015b; Livadiotis 2015c; 2016; Dialynas et al. 2018; Nicolaou & Livadiotis 2019),

$$\gamma = 1 + 2\, d_{\text{eff}}^{\;-1} \quad , \tag{39a}$$

that is,

$$\gamma = 1 + \frac{\alpha^2 + \frac{1}{2}}{(\alpha + \frac{1}{2})^2} \quad . \tag{39b}$$

Figure 3 plots the effective dimensionality and the adiabatic polytropic index with respect to the anisotropy. For anisotropy $\alpha$=1, the distribution is spherical, and the effective dimensionality equals the space dimensionality, $d_{\text{eff}} = 3$. As the anisotropy decreases for $\alpha$<1, reaching $\alpha$=0, the dimensionality decreases reaching $d_{\text{eff}} = 1$, describing a linear distribution (like a cigar); as the anisotropy increases for $\alpha$>1, reaching $\alpha \to \infty$, the dimensionality decreases reaching $d_{\text{eff}} = 2$, describing a “flat” distribution (like a pie). The effective dimensionality $d_{\text{eff}}$ has a local maximum at $\alpha$=1, while the corresponding adiabatic polytropic index $\gamma$ has a local minimum at $\alpha$=1.

We note that there are many space plasmas exhibiting positive correlations between density and temperature, with their most frequent polytropic index close to the value of the adiabatic process (e.g., Winterhalter et al. 1984; Tatrallyay et al. 1984; Totten et al. 1995; Newbury et al. 1997; Garcia 2001; Nicolaou et al. 2014a; Wang et al. 2015; 2016; Livadiotis & Desai 2016).

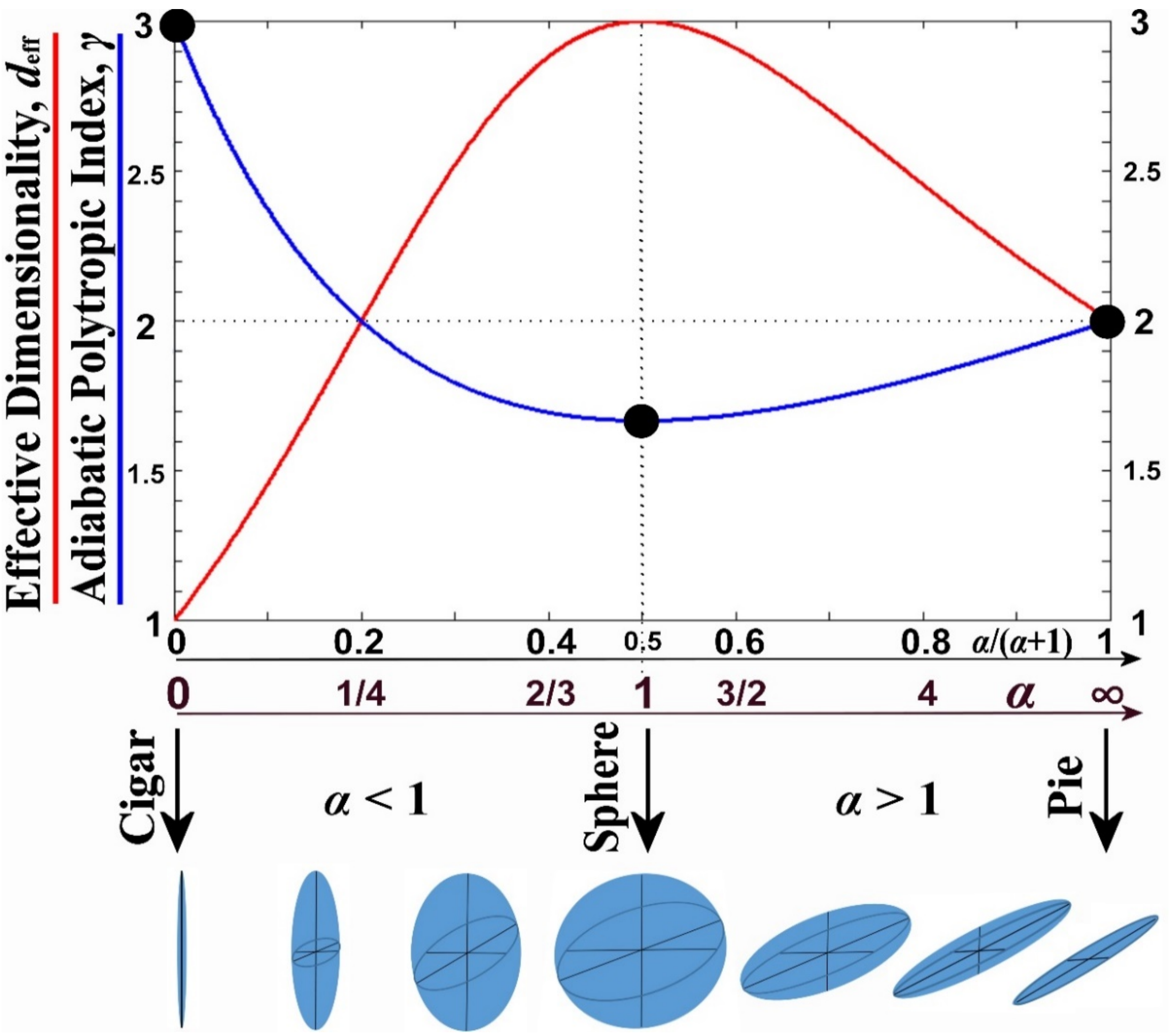

**Figure 3.** Effective dimensionality $d_{\text{eff}}$ (red) and the respective adiabatic polytropic index $\gamma$ (blue) as a function of the anisotropy $\alpha$, emphasizing the cases of $\alpha$=0, 1, ∞, corresponding to $d_{\text{eff}}$=1, 2, 3, and $\gamma$=3, 5/3, 2, where the velocity distribution takes the form of a cigar, sphere, pie.

## 4. Anisotropic kappa distributions with heterogeneously correlated variables

### *4.1. General*

Having established the importance of correlations among particles in the formulation of anisotropic kappa distributions, we now reverse the concept, in order to determine the variety of anisotropic expressions, which can be formulated based on the variety of particle correlations. In particular, here we develop and examine the various types of anisotropic kappa distributions, similar to the standard formula of Eq.(4a), which they can be derived from considering heterogeneous correlations among particles and their kinetic dof; first, we recognize the various types of homogeneous correlations (Subsection 6.2), and then, we proceed to the types of heterogeneous correlations (Subsection 6.3).

*4.2. Formulation of anisotropic kappa distributions based on homogeneous correlations*

The anisotropic kappa distribution shown in Eq.(4a) is a one-particle distribution that refers to systems with homogeneous correlation among the particle velocity components. The respective *N*-particle kappa distribution is given by

$$P\left(\{\vec{u}_i\}_{i=1}^{N};\theta_{\parallel},\theta_{\perp},\kappa_0;N\right)=(\pi\,\kappa_0)^{-\frac{3}{2}N}\cdot\frac{\Gamma(\kappa_0+1+\frac{3}{2}N)}{\Gamma(\kappa_0+1)}\cdot\theta_{\parallel}^{-N}\theta_{\perp}^{-2N}$$
$$\times\left[1+\frac{1}{\kappa_0}\cdot\left(\frac{1}{\theta_{\parallel}^{2}}\sum_{i=1}^{N}u_{\parallel i}^{\;2}+\frac{1}{\theta_{\perp}^{\;2}}\sum_{i=1}^{N}u_{\perp i}^{\;2}\right)\right]^{-\kappa_0-1-\frac{3}{2}N}\,, \tag{40a}$$

with normalization

$$\int_{-\infty}^{+\infty}\cdots(3N\,\text{integrals})\cdots\int_{-\infty}^{+\infty}P\left(\{\vec{u}_i\}_{i=1}^{N};\theta_{\parallel},\theta_{\perp},\kappa_0;N\right)d\vec{u}_1\ldots d\vec{u}_N\,=\,1\,, \tag{40b}$$

where $d\vec{u}_i\equiv du_{\parallel i}du_{\perp x_i}du_{\perp y_i}$. We recall the usage of the invariant kappa index $\kappa_0$ in multi-dimensional distributions (as explained in Subsections 2.2 and 2.3).

In the case where the two perpendicular velocity components are characterized by different temperature-like components, the anisotropic kappa distribution is written as

$$P\left(\{\vec{u}_i\}_{i=1}^{N};\theta_{\parallel},\theta_{\perp x},\theta_{\perp y},\kappa_0;N\right)=(\pi\,\kappa_0)^{-\frac{3}{2}N}\cdot\frac{\Gamma(\kappa_0+1+\frac{3}{2}N)}{\Gamma(\kappa_0+1)}\cdot\theta_{\parallel}^{-N}\theta_{\perp x}^{\;-N}\theta_{\perp y}^{\;-N}$$
$$\times\left[1+\frac{1}{\kappa_0}\cdot\left(\frac{1}{\theta_{\parallel}^{2}}\sum_{i=1}^{N}u_{\parallel i}^{2}+\frac{1}{\theta_{\perp x}^{2}}\sum_{i=1}^{N}u_{\perp xi}^{2}+\frac{1}{\theta_{\perp y}^{2}}\sum_{i=1}^{N}u_{\perp yi}^{2}\right)\right]^{-\kappa_0-1-\frac{3}{2}N}\,, \tag{41}$$

with normalization similar to that in Eq.(40b).

The base of the kappa distributions includes those velocity components that are characterized by the same correlation. Therefore, the correlation among the particles' parallel velocity components requires the inclusion of these components into the "base" of kappa distribution:

$$\{u_{\parallel_i}\xleftrightarrow{\kappa_0}u_{\parallel_j}\}:\quad u_{\parallel_1}\xleftrightarrow{\kappa_0}u_{\parallel_2}\xleftrightarrow{\kappa_0}\cdots\xleftrightarrow{\kappa_0}u_{\parallel_N}$$
$$\Updownarrow$$
$$P\left(\{u_{\parallel i}\}_{i=1}^{N};\theta_{\parallel},\cdots,\kappa_0;N\right)\sim\left[1+\frac{1}{\kappa_0}\cdot\left(\cdots+\frac{1}{\theta_{\parallel}^{2}}\sum_{i=1}^{N}u_{\parallel i}^{\;2}+\cdots\right)\right]^{-\kappa_0-1-\frac{1}{2}N-\cdots}\,, \tag{42}$$

where the existence of correlation is noted with "↔". We recall that the kappa index is defined by the correlation among particles; thus, the existence and notion of a correlation is given by the symbol of kappa index, $\kappa_0$ (or the 3D kappa index, here noted by $\kappa$).

Similarly, the existence of correlation among the particles' perpendicular velocity components requires the inclusion of the parallel components into the base of the kappa distribution:

$$
\begin{gathered}
\{u_{\perp xi} \overset{\kappa_0}{\longleftrightarrow} u_{\perp xj}\}: \quad u_{\perp x1} \overset{\kappa_0}{\longleftrightarrow} u_{\perp x2} \overset{\kappa_0}{\longleftrightarrow} \cdots \overset{\kappa_0}{\longleftrightarrow} u_{\perp xN} \ , \\
\Updownarrow \\
\left[1+\frac{1}{\kappa_0}\cdot\left(\cdots+\frac{1}{\theta_{\perp x}^2}\sum_{i=1}^{N} u_{\perp xi}^2+\cdots\right)\right]^{-\kappa_0-1-\frac{1}{2}N-\cdots} ,
\end{gathered} \tag{43}
$$

and

$$
\begin{gathered}
\{u_{\perp yi} \overset{\kappa_0}{\longleftrightarrow} u_{\perp yj}\}: \quad u_{\perp y1} \overset{\kappa_0}{\longleftrightarrow} u_{\perp y2} \overset{\kappa_0}{\longleftrightarrow} \cdots \overset{\kappa_0}{\longleftrightarrow} u_{\perp yN} \ , \\
\Updownarrow \\
\left[1+\frac{1}{\kappa_0}\cdot\left(\cdots+\frac{1}{\theta_{\perp y}^2}\sum_{i=1}^{N} u_{\perp yi}^2+\cdots\right)\right]^{-\kappa_0-1-\frac{1}{2}N-\cdots} .
\end{gathered} \tag{44}
$$

Furthermore, the correlation among the particles' parallel and perpendicular velocity components requires the inclusion of both the perpendicular and the parallel components into the base of the kappa distribution, i.e.,

$$
\begin{gathered}
\{u_{\|_i} \overset{\kappa_0}{\longleftrightarrow} u_{\|_j}\} \ \overset{\kappa_0}{\longleftrightarrow} \ \{u_{\perp xi} \overset{\kappa_0}{\longleftrightarrow} u_{\perp xj}\} \ , \\
\Updownarrow \\
P\left(\{u_{\|i}, u_{\perp xi}\}_{i=1}^{N}; N\right) \sim \left[1+\frac{1}{\kappa_0}\cdot\left(\frac{1}{\theta_{\|}^2}\sum_{i=1}^{N} u_{\|i}^2+\frac{1}{\theta_{\perp x}^2}\sum_{i=1}^{N} u_{\perp xi}^2+\cdots\right)\right]^{-\kappa_0-1-N-\cdots} ,
\end{gathered} \tag{45}
$$

and

$$
\begin{gathered}
\{u_{\|i} \overset{\kappa_0}{\longleftrightarrow} u_{\|j}\} \ \overset{\kappa_0}{\longleftrightarrow} \ \{u_{\perp xi} \overset{\kappa_0}{\longleftrightarrow} u_{\perp xj}\} \overset{\kappa_0}{\longleftrightarrow} \ \{u_{\perp yi} \overset{\kappa_0}{\longleftrightarrow} u_{\perp yj}\} \ , \\
\Updownarrow \\
P\left(\{u_{\|i}, u_{\perp xi}, u_{\perp yi}\}_{i=1}^{N}\right) \sim \left[1+\frac{1}{\kappa_0}\cdot\left(\frac{1}{\theta_{\|}^2}\sum_{i=1}^{N} u_{\|i}^2+\frac{1}{\theta_{\perp x}^2}\sum_{i=1}^{N} u_{\perp xi}^2+\frac{1}{\theta_{\perp y}^2}\sum_{i=1}^{N} u_{\perp yi}^2\right)\right]^{-\kappa_0-1-\frac{3}{2}N} .
\end{gathered} \tag{46}
$$

Since the correlation among alike particles' velocity components equals the correlation between different velocity components, the particle system is characterized by *homogeneous* correlations among particles' velocity components; symbolically, all correlations are characterized by the same kappa index, $\kappa_0$. We note that the one-particle kappa distribution, which is convenient and frequently used due to their simplicity, should not be interpreted as characterizing a case where the particles are not correlated to each other. On the contrary! The one-particle kappa distribution is simply a convenient mathematical tool, but physically an artifact; the correlations among the same velocity components of all the correlated particles, can be ignored when using the convenient one-particle kappa distribution, but still is implied through the concept of kappa index; (see Gravanis et al. 2020.)

Next, we examine the case of particle systems with heterogeneous correlations among their velocity components.

*4.3. Anisotropic distributions with heterogeneously correlated velocity components*

We consider the simplest case of unequal correlations among different velocity components. This is the case where the three dof, the parallel and the two perpendicular directions, are independent of each other, while the correlation among alike particles' dof is kept the same, i.e.,

$$\{u_{\|i} \overset{\kappa}{\longleftrightarrow} u_{\|j}\} \overset{\infty}{\longleftrightarrow} \{u_{\perp xi} \overset{\kappa}{\longleftrightarrow} u_{\perp xj}\} \overset{\infty}{\longleftrightarrow} \{u_{\perp yi} \overset{\kappa}{\longleftrightarrow} u_{\perp yj}\} \ , \tag{47a}$$

where the infinity notation among different dof means $\kappa \to \infty$, that is, no correlated or independent velocity components; for instance, Eq.(47a) reads that the $x$- and $y$- perpendicular components, as well as the parallel ($z$-) component, are all independent of each other. This simple type of heterogeneously correlated velocity components defines the distribution:

$$\begin{aligned} P(u_\|, \vec{u}_\perp; \theta_\|, \theta_\perp, \kappa) &= [\pi(\kappa - \tfrac{3}{2})]^{-\frac{3}{2}} \cdot \frac{\Gamma(\kappa)^3}{\Gamma(\kappa - \frac{1}{2})^3} \cdot \theta_\|^{-1} \theta_\perp^{-2} \\ &\times \left(1 + \frac{1}{\kappa - \frac{3}{2}} \cdot \frac{u_\|^2}{\theta_\|^2}\right)^{-\kappa} \cdot \left(1 + \frac{1}{\kappa - \frac{3}{2}} \cdot \frac{u_{\perp x}^2}{\theta_\perp^2}\right)^{-\kappa} \cdot \left(1 + \frac{1}{\kappa - \frac{3}{2}} \cdot \frac{u_{\perp y}^2}{\theta_\perp^2}\right)^{-\kappa} , \end{aligned} \tag{47b}$$

with normalization

$$\int_{-\infty}^{+\infty} \int_{-\infty}^{+\infty} \int_{-\infty}^{+\infty} P(u_\|, \vec{u}_\perp; \theta_\|, \theta_\perp, \kappa) \, du_\| du_{\perp x} du_{\perp y} = 1 \ . \tag{47c}$$

Another type of anisotropic distribution, but with heterogeneous correlations, is to have different correlations (or kappa indices) characterizing each dof, i.e.,

$$\{u_{xi} \overset{\kappa_x}{\longleftrightarrow} u_{xi}\} \overset{\infty}{\longleftrightarrow} \{u_{yi} \overset{\kappa_y}{\longleftrightarrow} u_{yj}\} \overset{\infty}{\longleftrightarrow} \{u_{zi} \overset{\kappa_z}{\longleftrightarrow} u_{zj}\} \ , \tag{48a}$$

that defines the distribution

$$\begin{aligned} P(\vec{u}; \vec{\theta}, \vec{\kappa}) &= \pi^{-\frac{3}{2}} \cdot \frac{(\kappa_x - \frac{3}{2})^{-\frac{1}{2}} \Gamma(\kappa_x)}{\Gamma(\kappa_x - \frac{1}{2})} \cdot \frac{(\kappa_y - \frac{3}{2})^{-\frac{1}{2}} \Gamma(\kappa_y)}{\Gamma(\kappa_y - \frac{1}{2})} \cdot \frac{(\kappa_z - \frac{3}{2})^{-\frac{1}{2}} \Gamma(\kappa_z)}{\Gamma(\kappa_z - \frac{1}{2})} \cdot \theta_x^{-1} \theta_y^{-1} \theta_z^{-1} \\ &\times \left(1 + \frac{1}{\kappa_x - \frac{3}{2}} \cdot \frac{u_x^2}{\theta_x^2}\right)^{-\kappa_x} \cdot \left(1 + \frac{1}{\kappa_y - \frac{3}{2}} \cdot \frac{u_y^2}{\theta_y^2}\right)^{-\kappa_y} \cdot \left(1 + \frac{1}{\kappa_z - \frac{3}{2}} \cdot \frac{u_z^2}{\theta_z^2}\right)^{-\kappa_z} , \end{aligned} \tag{48b}$$

with normalization

$$\int_{-\infty}^{+\infty} \int_{-\infty}^{+\infty} \int_{-\infty}^{+\infty} P(\vec{u}; \vec{\theta}, \vec{\kappa}) \, du_x du_y du_z = 1 \ , \tag{48c}$$

where we set the triads (not vectors) $\vec{\theta} \equiv (\theta_x, \theta_y, \theta_z)$ and $\vec{\kappa} \equiv (\kappa_x, \kappa_y, \kappa_z)$.

A more complicate case is when the heterogeneous correlations vary among different pairs of components, e.g., the two perpendicular components $x$ and $y$ are correlated with a common kappa index, while both of them are independent with the parallel component ($\kappa \to \infty$), that is,

$$\{u_{\|i} \overset{\kappa}{\longleftrightarrow} u_{\|j}\} \overset{\infty}{\longleftrightarrow} \{u_{\perp xi} \overset{\kappa}{\longleftrightarrow} u_{\perp xj}\} \overset{\kappa}{\longleftrightarrow} \{u_{\perp yi} \overset{\kappa}{\longleftrightarrow} u_{\perp yj}\} \ , \tag{49a}$$

that defines the distribution

$$P(u_{\|},\vec{u}_{\perp};\theta_{\|},\theta_{\perp},\kappa)=[\pi\,(\kappa-\tfrac{3}{2})]^{-\frac{3}{2}}\cdot\frac{\Gamma(\kappa)(\kappa-\frac{1}{2})}{\Gamma(\kappa-\frac{1}{2})}\cdot\theta_{\|}^{-1}\theta_{\perp}^{-2}$$

$$\times\left(1+\frac{1}{\kappa-\frac{3}{2}}\cdot\frac{u_{\|}^{2}}{\theta_{\|}^{2}}\right)^{-\kappa}\cdot\left(1+\frac{1}{\kappa-\frac{3}{2}}\cdot\frac{u_{\perp x}^{2}+u_{\perp y}^{2}}{\theta_{\perp}^{2}}\right)^{-\kappa-\frac{1}{2}}\ , \tag{49b}$$

with normalization

$$\int_{0}^{\infty}\int_{-\infty}^{+\infty}P(u_{\|},u_{\perp};\theta_{\|},\theta_{\perp},\kappa)\,du_{\|}\,2\pi u_{\perp}du_{\perp}\,=\,1\ , \tag{49c}$$

(e.g., for applications, see: Summers & Thorne 1991, Dos Santos et al. 2015).

The above anisotropic distributions constitute trivial types of heterogeneous correlations among the particles' velocity components. This is because the heterogeneous correlation, i.e., the correlation connecting different dof, is either zero or equal to the correlation connecting alike dof; namely, the corresponding kappa value is either infinity or equal to the kappa index of the distribution. Next, we will present the general case that describes the non-trivial heterogeneous correlations; but first, we have to find a way to express the correlations in terms of probability distributions; for this, we focus on the expression of the multi-dimensional kappa probability distribution in terms of its marginal probability distributions.

*4.4. Correlation coefficients for anisotropic distributions subject to heterogeneous correlations*

Let the anisotropic kappa distribution in Eq.(47b) with heterogeneous correlation equal to zero, $\kappa^{\mathrm{int}}\rightarrow\infty$, i.e., the three velocity components are independent of each other; hence, the ensemble of all particles' components constitutes a system with one correlated component, that is, with effective dimensionality $d_{\mathrm{eff}}=1$; thus, the correlation coefficient is

$$\rho=\frac{\frac{1}{2}}{\kappa_{0}+\frac{1}{2}}=\frac{\frac{1}{2}}{\kappa_{3}-1}=\frac{\frac{1}{2}}{\kappa_{1}}\ , \tag{50}$$

which is written in terms of the invariant kappa index $\kappa_0$, the typical 3D kappa index $\kappa_3$, or the 1D kappa index $\kappa_1$; (compare to Eq.(34)).

In the case of distributions with unequal heterogeneous correlations the derivations are more complicated and will be presented elsewhere. Here we just mention the case of distributions of $d$-D velocity, decomposed to 1D parallel component and a ($d$-1)-D perpendicular component. The correlation coefficient is

$$\rho=\frac{1+(d-1)^{2}\alpha^{2}}{\kappa_{0}+1+(\kappa_{0}+d-1)(d-1)\alpha^{2}}\ , \tag{51}$$

which can be written again in its standard formulation, as in Eq.(36),

$$\rho=\frac{\frac{1}{2}d_{\mathrm{eff}}}{\kappa_{0}+\frac{1}{2}d_{\mathrm{eff}}}\ \text{ or }\ \rho=\frac{\frac{1}{2}d_{\mathrm{eff}}}{(\kappa_{3}-\frac{3}{2})+\frac{1}{2}d_{\mathrm{eff}}}\ , \tag{52}$$

where the effective dimensionality and adiabatic polytropic index are respectively given by

$$d_{\mathrm{eff}}(\alpha,d)=\frac{1+(d-1)^2\alpha^2}{1+(d-1)\alpha^2} \ . \tag{53}$$

$$\gamma(\alpha,d)=1+2\,d_{\mathrm{eff}}(\alpha,d)^{-1}=1+2\cdot\frac{1+(d-1)\alpha^2}{1+(d-1)^2\alpha^2} \ . \tag{54}$$

## 5. Generalized anisotropic kappa distributions based on heterogeneity

### *5.1. General*

Previously, we have shown the types of homogeneous and heterogeneous correlations among the particle velocity components. We examined the typical cases, where the velocity components of each particle are either (a) homogeneously correlated, that is, the correlations from all components are characterized by the same kappa index, or (b) uncorrelated, that is, the correlations between different components are characterized by kappa index equal to infinity. We now develop the general case, where the velocity components can be heterogeneously correlated with an arbitrary heterogeneity, namely, the correlations between different components can be characterized by any kappa index.

The generalization of anisotropic kappa distributions takes place within the framework of Tsallis nonextensive statistical mechanics (e.g., 2009; Tsallis et al. 1998). This is based on a natural generalization of entropy, the so-called *q*-entropy (or Tsallis entropy, e.g., see: Tsallis 1988, Livadiotis 2018a;b), while the statistical origin of kappa distributions is connected with nonextensive statistical mechanics, where the associated entropy is maximized under the constraints of canonical ensemble. Certainly, Tsallis nonextensive statistical mechanics is well known in the community of statistical (Tsallis 2009), space (Livadiotis 2017) and plasma (Yoon 2019) physics. On the other hand, other similar promising theories exist, but they need to be further developed, e.g., Superstatistics (Beck & Cohen 2001), Kaniadakis entropy and its maximization (Kaniadakis 2001; Macedo-Filho et al. 2013).

As shown in Livadiotis & McComas (2009; 2013), the kappa distribution coincides exactly with the *q*-exponential distribution, the one that emerges from the maximization of *q*-entropy, under the trivial transformation of their indices $\kappa=1/(q-1)$. The equivalence between the formulations of kappa and *q*-exponential distributions becomes obvious in the case of particle systems with continuous energy density. Surely, there other cases, such as systems with quantum energy states, where the distribution resulting from the maximization of entropy is not given by a *q*-exponential or kappa distribution. Moreover, it has to be noted that kappa distributions characterize systems residing in stationary states, and they are consistent with the concept of thermal equilibrium (Livadiotis 2018a). While there are various mechanisms responsible for generating kappa distributions (Livadiotis et al. 2018; 2019b), it is their consistency with thermal equilibrium that allows the existence of these distributions.

The concept of correlation is interwoven with the theory of kappa distributions and nonextensive statistical mechanics. Let two sets A and B of discrete probability distributions be $\{p_i^A\}_{i=1}^W = p_1^A, p_2^A, \ldots, p_W^A$ and $\{p_j^B\}_{j=1}^W = p_1^B, p_2^B, \ldots, p_W^B$, as well as their 2D joint probability distribution, $\{p_{ij}^{A+B}\}_{j=1}^W$ $= p_{11}^{A+B}, \ldots, p_{1W}^{A+B}, p_{21}^{A+B}, \ldots, p_{2W}^{A+B}, \ldots, p_{W1}^{A+B}, \ldots, p_{WW}^{A+B}$. The classical Boltzmann-Gibbs statistical mechanics requires that there is no correlation between the particles (Gibbs 1902), i.e., $p_{ij}^{A+B} = p_i^A \cdot p_j^B$, which leads to the additivity of Boltzmann's entropy, i.e., $S^{A+B} = S^A + S^B$ (Tsallis 2005). On the other hand, non-extensive Statistical Mechanics implies a certain type of correlation, that is $p_{ij}^{A+B} = g(p_i^A, p_j^B; a)$, where $g(x, y; a) \equiv [x^{-a} + y^{-a} - 1]^{-1/a}$. Then we can easily show that the Boltzmann's entropy is non-additive, where the additivity rule is also characterized by nonlinear coupling term $S^{A+B} = S^A + S^B - a \cdot S^A \cdot S^B$, but, the Tsallis entropy is still additive, under this type of correlation (Tsallis 2005, Livadiotis 2018a;b). The exponent can be easily related to the kappa index $\kappa \equiv 1/a$, while the above type of correlations will be used to construct the formulation of anisotropic kappa distributions with nontrivial heterogeneity among the particle velocity components.

### *5.2. Joint-probability distribution and its partition to marginal probability distributions*

Let the $f$-dimensional kappa distribution describing $f = N \cdot d$ velocity components, $\{u_i\}_{i=1}^f = \{u_1, \cdots, u_f\}$, according to the notation in Subsection §2.3, that is,

$$P(u_1, u_2, \cdots, u_f; \theta, \kappa_0; f) = (\pi\kappa_0)^{-\frac{1}{2}f} \cdot \frac{\Gamma(\kappa_0 + 1 + \frac{1}{2}f)}{\Gamma(\kappa_0 + 1)} \cdot \theta^{-f} \cdot \left[1 + \frac{1}{\kappa_0} \cdot \frac{u_1^2 + u_2^2 + \cdots + u_f^2}{\theta^2}\right]^{-\kappa_0 - 1 - \frac{1}{2}f} , \quad (55)$$

where we recall again the requirement of using the invariant kappa index $\kappa_0$ in the formulation of multi-dimensional distributions.

Furthermore, we use the variables of the total kinetic energy and the kinetic energy of the $i^{\text{th}}$ component, which are, respectively,

$$\varepsilon_{\text{tot}\,f} = \tfrac{1}{2} m \sum_{i=1}^{f} u_i^2 = \tfrac{1}{2} m \sum_{i=1}^{f} \varepsilon_i \text{ and } \varepsilon_i = \tfrac{1}{2} m u_i^2 . \quad (56)$$

Hence, we rewrite the distribution (55) as

$$P(\varepsilon_{\text{tot}\,f}) = P(0) \cdot \left(1 + \frac{1}{\kappa_0} \cdot \frac{\varepsilon_{\text{tot}\,f}}{k_{\text{B}} T}\right)^{-\kappa_0 - 1 - \frac{1}{2}f} , \quad (57)$$

while the one-particle distribution of any $i^{\text{th}}$ component is similarly written as

$$P(\varepsilon_i) = P(0) \cdot \left(1 + \frac{1}{\kappa_0} \cdot \frac{\varepsilon_i}{k_{\text{B}} T}\right)^{-\kappa_0 - 1 - \frac{1}{2}} , \; i: 1, \ldots, f. \quad (58)$$

Then, we obtain

$$\left[\frac{P(\varepsilon_i)}{P(0)}\right]^{-\frac{1}{\kappa_0-1-\frac{1}{2}}}-1=\frac{1}{\kappa_0 k_{\mathrm{B}}T}\cdot\varepsilon_i \ , \tag{59}$$

and

$$\left[\frac{P(\varepsilon_{\mathrm{tot}\,f})}{P(0)}\right]^{-\frac{1}{\kappa_0-1-\frac{1}{2}f}}-1=\frac{1}{\kappa_0 k_{\mathrm{B}}T}\cdot\varepsilon_{\mathrm{tot}\,f}=\frac{1}{\kappa_0 k_{\mathrm{B}}T}\cdot\sum_{i=1}^{f}\varepsilon_i=\sum_{i=1}^{f}\left\{\left[\frac{P(\varepsilon_i)}{P(0)}\right]^{-\frac{1}{\kappa_0-1-\frac{1}{2}}}-1\right\} . \tag{60}$$

Therefore, the partition of the joint probability distribution, $P(\varepsilon_{\mathrm{tot}\,f})$ or $P(u_1,u_2,\cdots,u_f)$, with its marginal probability distributions, $P(\varepsilon_i)$ or $P(u_i)$, is given by the relation

$$\left[\frac{P(\varepsilon_{\mathrm{tot}\,f})}{P(0)}\right]^{-\frac{1}{\kappa_0-1-\frac{1}{2}f}}-1=\sum_{i=1}^{f}\left\{\left[\frac{P(\varepsilon_i)}{P(0)}\right]^{-\frac{1}{\kappa_0-1-\frac{1}{2}}}-1\right\} , \tag{61a}$$

which generalizes the respective independence relation of Maxwell-Boltzmann distributions, corresponding to $\kappa_0\rightarrow\infty$,

$$\frac{P(\varepsilon_{\mathrm{tot}\,f})}{P(0)}=\prod_{i=1}^{f}\left[\frac{P(\varepsilon_i)}{P(0)}\right] . \tag{61b}$$

The finite value of the kappa index involved in the partition relation (61a) is the key-parameter setting the correlations of the joint probability distribution, $P(\varepsilon_{\mathrm{tot}\,f})$ or $P(u_1,u_2,\cdots,u_f)$. For an arbitrary kappa index, a unique correlation type is determined among the velocity components. These statistical correlations can be caused by existing, or even pre-existing, interactions among particles, which reserve the correlations, especially in the case of collisionless particle systems; the phenomenological kappa index is noted by $\kappa_0^{\mathrm{int}}$ (i.e., due to "interactions"), and Eq.(61a) leads to the partition relation

$$\left[\frac{P(\varepsilon_{\mathrm{tot}\,f})}{P(0)}\right]^{-\frac{1}{\kappa_0^{\mathrm{int}}-1-\frac{1}{2}f}}-1=\sum_{i=1}^{f}\left\{\left[\frac{P(\varepsilon_i)}{P(0)}\right]^{-\frac{1}{\kappa_0^{\mathrm{int}}-1-\frac{1}{2}}}-1\right\} , \tag{62a}$$

leading to the distribution

$$P(\varepsilon_{\mathrm{tot}\,f})=P(0)\cdot\left\{1+\sum_{i=1}^{f}\left\{\left[\frac{P(\varepsilon_i)}{P(0)}\right]^{-\frac{1}{\kappa_0^{\mathrm{int}}-1-\frac{1}{2}}}-1\right\}\right\}^{-\kappa_0^{\mathrm{int}}-1-\frac{1}{2}f} . \tag{62b}$$

Next, we see how the partition relation (62a) is used for developing the formulation of more complicated types of anisotropic distributions, covering all combinations of heterogeneous correlations among the particles' velocity components.

*5.3. Anisotropic distributions with nontrivial heterogeneity*

Let the kappa index $\kappa^{\text{int}}$ characterize the heterogeneous correlations between any two dof, either perpendicular or parallel velocity components, namely,

$$\{u_{xi} \overset{\kappa_x}{\longleftrightarrow} u_{xj}\} \overset{\kappa^{\text{int}}}{\longleftrightarrow} \{u_{yi} \overset{\kappa_y}{\longleftrightarrow} u_{yj}\} \overset{\kappa^{\text{int}}}{\longleftrightarrow} \{u_{zi} \overset{\kappa_z}{\longleftrightarrow} u_{zj}\} \ . \tag{63a}$$

This defines the distribution

$$P(\vec{u};\vec{\theta},\vec{\kappa}) \sim \left\{-2+\left(1+\frac{1}{\kappa_x-\frac{3}{2}}\cdot\frac{{u_x}^2}{{\theta_x}^2}\right)^{\frac{\kappa_x}{\kappa^{\text{int}}}}+\left(1+\frac{1}{\kappa_y-\frac{3}{2}}\cdot\frac{{u_y}^2}{{\theta_y}^2}\right)^{\frac{\kappa_y}{\kappa^{\text{int}}}}+\left(1+\frac{1}{\kappa_z-\frac{3}{2}}\cdot\frac{{u_z}^2}{{\theta_z}^2}\right)^{\frac{\kappa_z}{\kappa^{\text{int}}}}\right\}^{-\kappa^{\text{int}}-1} . \tag{63b}$$

A more general case concerns heterogeneous correlations with different kappa value between different velocity components. As a typical example, we consider the case where two different kappa values, $\kappa_1^{\text{int}}$ and $\kappa_2^{\text{int}}$, describe the heterogeneous correlations among different dof; in particular, $\kappa_1^{\text{int}}$ describes the correlations between components *x* and *y*, while $\kappa_2^{\text{int}}$ describes correlations between *x* or *y* components with the *z* component, i.e.,

$$\left\{\{u_{xi} \overset{\kappa_x}{\longleftrightarrow} u_{xj}\} \overset{\kappa_1^{\text{int}}}{\longleftrightarrow} \{u_{yi} \overset{\kappa_y}{\longleftrightarrow} u_{yj}\}\right\} \overset{\kappa_2^{\text{int}}}{\longleftrightarrow} \{u_{zi} \overset{\kappa_z}{\longleftrightarrow} u_{zj}\} \ , \tag{64a}$$

which defines the distribution

$$P(\vec{u};\vec{\theta},\vec{\kappa}) \sim \left\{-1+\left\{-1+\left(1+\frac{1}{\kappa_x-\frac{3}{2}}\cdot\frac{{u_x}^2}{{\theta_x}^2}\right)^{\frac{\kappa_x}{\kappa_1^{\text{int}}}}+\left(1+\frac{1}{\kappa_y-\frac{3}{2}}\cdot\frac{{u_y}^2}{{\theta_y}^2}\right)^{\frac{\kappa_y}{\kappa_1^{\text{int}}}}\right\}^{\frac{\kappa_1^{\text{int}}+\frac{1}{2}}{\kappa_2^{\text{int}}+\frac{1}{2}}}+\left(1+\frac{1}{\kappa_z-\frac{3}{2}}\cdot\frac{{u_z}^2}{{\theta_z}^2}\right)^{\frac{\kappa_z}{\kappa_2^{\text{int}}}}\right\}^{-\kappa_2^{\text{int}}-1} . \tag{64b}$$

This can be focused to the perpendicular/parallel notation, as follows:

$$\left\{\{u_{\perp xi} \overset{\kappa_{\perp x}}{\longleftrightarrow} u_{\perp xj}\} \overset{\kappa_{\perp}^{\text{int}}}{\longleftrightarrow} \{u_{\perp yi} \overset{\kappa_{\perp y}}{\longleftrightarrow} u_{\perp yj}\}\right\} \overset{\kappa_{\perp\|}^{\text{int}}}{\longleftrightarrow} \{u_{\| i} \overset{\kappa_{\|}}{\longleftrightarrow} u_{\| j}\} \ , \tag{65a}$$

and

$$P(\vec{u};\vec{\theta},\vec{\kappa}) \sim \left\{-1+\left\{-1+\left(1+\frac{1}{\kappa_{\perp x}-\frac{3}{2}}\cdot\frac{{u_{\perp x}}^2}{{\theta_{\perp x}}^2}\right)^{\frac{\kappa_{\perp x}}{\kappa_{\perp}^{\text{int}}}}+\left(1+\frac{1}{\kappa_{\perp y}-\frac{3}{2}}\cdot\frac{{u_{\perp y}}^2}{{\theta_{\perp y}}^2}\right)^{\frac{\kappa_{\perp y}}{\kappa_{\perp}^{\text{int}}}}\right\}^{\frac{\kappa_{\perp}^{\text{int}}+\frac{1}{2}}{\kappa_{\perp\|}^{\text{int}}+\frac{1}{2}}}+\left(1+\frac{1}{\kappa_{\|}-\frac{3}{2}}\cdot\frac{{u_{\|}}^2}{{\theta_{\|}}^2}\right)^{\frac{\kappa_{\|}}{\kappa_{\perp\|}^{\text{int}}}}\right\}^{-\kappa_{\perp\|}^{\text{int}}-1} . \tag{65b}$$

A more interesting case is when $\kappa_{\perp x} = \kappa_{\perp y} = \kappa_{\perp} = \kappa^{\text{int}}$, i.e., the correlations become

$$\left\{\{u_{\perp xi} \overset{\kappa_\perp}{\longleftrightarrow} u_{\perp xj}\} \overset{\kappa_\perp}{\longleftrightarrow} \{u_{\perp yi} \overset{\kappa_\perp}{\longleftrightarrow} u_{\perp yj}\}\right\} \overset{\kappa_{\perp\parallel}^{\rm int}}{\longleftrightarrow} \{u_{\parallel i} \overset{\kappa_\parallel}{\longleftrightarrow} u_{\parallel j}\} \ , \tag{66a}$$

with distribution

$$P(u_\parallel, u_\perp; \theta_\parallel, \theta_\perp, \kappa_\parallel, \kappa_\perp, \kappa^{\rm int}) \sim \left\{-1 + \left(1 + \frac{1}{\kappa_\perp - \frac{3}{2}} \cdot \frac{u_\perp^{\,2}}{\theta_\perp^{\,2}}\right)^{\frac{\kappa_\perp + \frac{1}{2}}{\kappa^{\rm int} + \frac{1}{2}}} + \left(1 + \frac{1}{\kappa_\parallel - \frac{3}{2}} \cdot \frac{u_\parallel^{\,2}}{\theta_\parallel^{\,2}}\right)^{\frac{\kappa_\parallel}{\kappa^{\rm int}}}\right\}^{-\kappa^{\rm int} - 1} . \tag{66b}$$

This is even more simplified when $\kappa_\parallel = \kappa_\perp = \kappa$, i.e., the correlations are written as

$$\left\{\{u_{\perp xi} \overset{\kappa}{\longleftrightarrow} u_{\perp xj}\} \overset{\kappa}{\longleftrightarrow} \{u_{\perp yi} \overset{\kappa}{\longleftrightarrow} u_{\perp yj}\}\right\} \overset{\kappa^{\rm int}}{\longleftrightarrow} \{u_{\parallel i} \overset{\kappa}{\longleftrightarrow} u_{\parallel j}\} \ , \tag{67a}$$

with distribution

$$P(u_\parallel, u_\perp; \theta_\parallel, \theta_\perp, \kappa, \kappa^{\rm int}) \sim \left\{-1 + \left(1 + \frac{1}{\kappa - \frac{3}{2}} \cdot \frac{u_\perp^{\,2}}{\theta_\perp^{\,2}}\right)^{\frac{\kappa + \frac{1}{2}}{\kappa^{\rm int} + \frac{1}{2}}} + \left(1 + \frac{1}{\kappa - \frac{3}{2}} \cdot \frac{u_\parallel^{\,2}}{\theta_\parallel^{\,2}}\right)^{\frac{\kappa}{\kappa^{\rm int}}}\right\}^{-\kappa^{\rm int} - 1} . \tag{67b}$$

The normalization constant of these distributions does not exist in a closed-form expression. It is given by $\mathrm{C}(\kappa^{\rm int}, \kappa) \cdot \alpha^{-1} [\frac{1}{3}(1+2\alpha)]^{\frac{3}{2}} \theta^{-3}$, where

$$\mathrm{C}(\kappa^{\rm int}, \kappa)^{-1} = \int_0^\infty \int_{-\infty}^{+\infty} \left[-1 + \left(1 + \frac{x_\perp^{\,2}}{\kappa - \frac{3}{2}}\right)^{\frac{\kappa + \frac{1}{2}}{\kappa^{\rm int} + \frac{1}{2}}} + \left(1 + \frac{x_\parallel^{\,2}}{\kappa - \frac{3}{2}}\right)^{\frac{\kappa}{\kappa^{\rm int}}}\right]^{-\kappa^{\rm int} - 1} dx_\parallel \, 2\pi x_\perp dx_\perp \tag{68a}$$

In Appendix, we show that

$$\mathrm{C}(\kappa^{\rm int}, \kappa) = \pi^{-1} \cdot \frac{\kappa}{\kappa^{\rm int}} \cdot \frac{\kappa^{\rm int}(\kappa - \frac{1}{2}) + \kappa}{(\kappa - \frac{3}{2})^{\frac{3}{2}} (\kappa^{\rm int} + \frac{1}{2})}$$
$$\times \left[\sum_{m=0}^{\infty} \frac{(2m)!}{4^m (m!)^2 [\kappa^{\rm int} + 1 + \frac{\kappa^{\rm int}}{\kappa}(m - \frac{1}{2})]} \cdot {}_3F_2\left(\kappa^{\rm int} + 1, 1, 1; \kappa^{\rm int} + 2 - \frac{\kappa^{\rm int} + \frac{1}{2}}{\kappa + \frac{1}{2}}, \kappa^{\rm int} + 2 + \frac{\kappa^{\rm int}}{\kappa}(m - \tfrac{1}{2}); 1\right)\right]^{-1} , \tag{68b}$$

or

$$\mathrm{C}(\kappa^{\rm int}, \kappa) = \pi^{-1} \cdot \frac{\kappa[\kappa^{\rm int}(\kappa - \frac{1}{2}) + \kappa]}{(\kappa - \frac{3}{2})^{\frac{3}{2}} (\kappa^{\rm int} + \frac{1}{2})} \cdot \frac{(\kappa^{\rm int} - 1)!}{[\kappa^{\rm int} + 1 - \frac{\kappa^{\rm int} + \frac{1}{2}}{\kappa + \frac{1}{2}}]!}$$
$$\times \left[\sum_{m=0}^{\infty} \frac{(2m)! [\kappa^{\rm int} + \frac{\kappa^{\rm int}}{\kappa}(m - \frac{1}{2})]!}{4^m (m!)^2} \cdot \sum_{n=0}^{\infty} \frac{(\kappa^{\rm int} + n)! n!}{[\kappa^{\rm int} + 1 - \frac{\kappa^{\rm int} + \frac{1}{2}}{\kappa + \frac{1}{2}} + n]! \, [\kappa^{\rm int} + 1 + \frac{\kappa^{\rm int}}{\kappa}(m - \frac{1}{2}) + n]!}\right]^{-1} . \tag{68c}$$

Finally, it may be useful in certain cases to have the distribution expanded in terms of $(\kappa^{\rm int} - \kappa) \equiv \delta << 1$, i.e.,

$$P(u_{\|},u_{\perp};\theta_{\|},\theta_{\perp},\kappa,\kappa^{\text{int}}) \sim \left[1+\frac{1}{\kappa-\frac{3}{2}}\cdot\left(\frac{u_{\perp}^{2}}{\theta_{\perp}^{2}}+\frac{u_{\|}^{2}}{\theta_{\|}^{2}}\right)\right]^{-\kappa-1} + \delta\cdot\Delta P(u_{\|},u_{\perp};\theta_{\|},\theta_{\perp},\kappa) \ , \tag{69a}$$

where

$$\Delta P(u_{\|},u_{\perp};\theta_{\|},\theta_{\perp},\kappa) = (\kappa+1)\cdot\left[1+\frac{1}{\kappa-\frac{3}{2}}\cdot\left(\frac{u_{\perp}^{2}}{\theta_{\perp}^{2}}+\frac{u_{\|}^{2}}{\theta_{\|}^{2}}\right)\right]^{-\kappa-2}$$
$$\times\left\{\begin{array}{l} \frac{1}{\kappa+\frac{1}{2}}\left(1+\frac{1}{\kappa-\frac{3}{2}}\cdot\frac{u_{\perp}^{2}}{\theta_{\perp}^{2}}\right)\ln\left(1+\frac{1}{\kappa-\frac{3}{2}}\cdot\frac{u_{\perp}^{2}}{\theta_{\perp}^{2}}\right) \\ +\frac{1}{\kappa}\left(1+\frac{1}{\kappa-\frac{3}{2}}\cdot\frac{u_{\|}^{2}}{\theta_{\|}^{2}}\right)\ln\left(1+\frac{1}{\kappa-\frac{3}{2}}\cdot\frac{u_{\|}^{2}}{\theta_{\|}^{2}}\right) \\ -\frac{1}{\kappa+1}\left[1+\frac{1}{\kappa-\frac{3}{2}}\cdot\left(\frac{u_{\perp}^{2}}{\theta_{\perp}^{2}}+\frac{u_{\|}^{2}}{\theta_{\|}^{2}}\right)\right]\ln\left[1+\frac{1}{\kappa-\frac{3}{2}}\cdot\left(\frac{u_{\perp}^{2}}{\theta_{\perp}^{2}}+\frac{u_{\|}^{2}}{\theta_{\|}^{2}}\right)\right] \end{array}\right\} . \tag{69b}$$

*5.4. Summary of anisotropic distributions formulae*

Table 2 summarizes the main formulae of anisotropic kappa distributions, as determined from the type of homogeneous/heterogeneous correlations.

The main three and more useful anisotropic distributions are the following:

- Heterogeneous correlation equal to the correlation of each component, $\kappa^{\text{int}}=\kappa$,

$$P(\vec{u};\theta,\alpha,\kappa;\kappa^{\text{int}}=\kappa) = \pi^{-\frac{3}{2}}\cdot(\kappa-\tfrac{3}{2})^{-\frac{3}{2}}\frac{\Gamma(\kappa+1)}{\Gamma(\kappa-\frac{1}{2})}$$
$$\times\,\alpha^{-1}[\tfrac{1}{3}(1+2\alpha)]^{\frac{3}{2}}\theta^{-3}\cdot\left[1+\frac{1}{\kappa-\frac{3}{2}}\cdot\frac{1+2\alpha}{3\alpha\theta^{2}}\cdot(\alpha u_{\|}^{2}+u_{\perp}^{2})\right]^{-\kappa-1} . \tag{70a}$$

- Heterogeneous correlation (between perpendicular and parallel components) equal to zero, $\kappa^{\text{int}}\rightarrow\infty$,

$$P(\vec{u};\theta,\alpha,\kappa;\kappa^{\text{int}}\rightarrow\infty) = [\pi\,(\kappa-\tfrac{3}{2})]^{-\frac{3}{2}}\cdot\frac{\Gamma(\kappa)(\kappa-\frac{1}{2})}{\Gamma(\kappa-\frac{1}{2})}\cdot\alpha^{-1}[\tfrac{1}{3}(1+2\alpha)]^{\frac{3}{2}}\theta^{-3}$$
$$\times\left(1+\frac{1}{\kappa-\frac{3}{2}}\cdot\frac{1+2\alpha}{3\alpha\theta^{2}}\cdot u_{\perp}^{2}\right)^{-\kappa-\frac{1}{2}}\cdot\left(1+\frac{1}{\kappa-\frac{3}{2}}\cdot\frac{1+2\alpha}{3\theta^{2}}\cdot u_{\|}^{2}\right)^{-\kappa} . \tag{70b}$$

- Heterogeneous correlation between perpendicular and parallel components) has an arbitrary value, $\kappa^{\text{int}}<\infty$,

$$P(\vec{u};\theta,\alpha,\kappa;\kappa^{\text{int}}) = \mathrm{C}(\kappa^{\text{int}},\kappa)\cdot\alpha^{-1}[\tfrac{1}{3}(1+2\alpha)]^{\frac{3}{2}}\theta^{-3}$$
$$\times\left[-1+\left(1+\frac{1}{\kappa-\frac{3}{2}}\cdot\frac{1+2\alpha}{3\alpha\theta^{2}}\cdot u_{\perp}^{2}\right)^{\frac{\kappa+\frac{1}{2}}{\kappa^{\text{int}}+\frac{1}{2}}}+\left(1+\frac{1}{\kappa-\frac{3}{2}}\cdot\frac{1+2\alpha}{3\theta^{2}}\cdot u_{\|}^{2}\right)^{\frac{\kappa}{\kappa^{\text{int}}}}\right]^{-\kappa^{\text{int}}-1} . \tag{70c}$$

The developed distribution in Eq.(70c) mediates the main two types of anisotropic kappa distributions, where the first in Eq.(70a) considers equal correlations among particles velocity components, while the second in Eq.(70b) considers zero correlation among different velocity components. (Appendix A helps on the normalization of those distributions). Figure 4 plots the distribution (70b), while Figures 5 and 6 plot the distribution (70c), for $\kappa^{\text{int}}$= 3 and 1.5, respectively; the plotted velocity components are normalized to the thermal speed, $\theta$.

**Table 2.** Main formulae of anisotropic kappa distributions in association with the correlation type

| | Correlation type * | Distribution $P(\vec{u})$ |
|---|---|---|
| 1 | $\{u_{\parallel i} \overset{\kappa}{\longleftrightarrow} u_{\parallel j}\} \overset{\kappa}{\longleftrightarrow} \{u_{\perp xi} \overset{\kappa}{\longleftrightarrow} u_{\perp xj}\} \overset{\kappa}{\longleftrightarrow} \{u_{\perp yi} \overset{\kappa}{\longleftrightarrow} u_{\perp yj}\}$<br>Homogeneous correlations | $\pi^{-\frac{3}{2}}(\kappa-\tfrac{3}{2})^{-\frac{3}{2}}\cdot\frac{\Gamma(\kappa+1)}{\Gamma(\kappa-\frac{1}{2})}\cdot\theta_{\parallel}^{-1}\theta_{\perp}^{-2}\cdot\left[1+\frac{1}{\kappa-\frac{3}{2}}\cdot\left(\frac{u_{\parallel}^{2}}{\theta_{\parallel}^{2}}+\frac{u_{\perp}^{2}}{\theta_{\perp}^{2}}\right)\right]^{-\kappa-1}$ |
| 2a | $\{u_{\parallel i} \overset{\kappa}{\longleftrightarrow} u_{\parallel j}\} \overset{\infty}{\longleftrightarrow} \{u_{\perp xi} \overset{\kappa}{\longleftrightarrow} u_{\perp xj}\} \overset{\infty}{\longleftrightarrow} \{u_{\perp yi} \overset{\kappa}{\longleftrightarrow} u_{\perp yj}\}$<br>Heterogeneous correlations.<br>Alike dof: $\kappa$; Different dof: $\kappa\to\infty$. | $\pi^{-\frac{3}{2}}\cdot\frac{(\kappa-\frac{3}{2})^{-\frac{3}{2}}\Gamma(\kappa)^{3}}{\Gamma(\kappa-\frac{1}{2})^{3}}\cdot\theta_{\parallel}^{-1}\theta_{\perp}^{-2}\left(1+\frac{1}{\kappa-\frac{3}{2}}\cdot\frac{u_{\parallel}^{2}}{\theta_{\parallel}^{2}}\right)^{-\kappa}\left(1+\frac{1}{\kappa-\frac{3}{2}}\cdot\frac{u_{\perp x}^{2}}{\theta_{\perp}^{2}}\right)^{-\kappa}\left(1+\frac{1}{\kappa-\frac{3}{2}}\cdot\frac{u_{\perp y}^{2}}{\theta_{\perp}^{2}}\right)^{-\kappa}$ |
| 2b | $\{u_{xi} \overset{\kappa_x}{\longleftrightarrow} u_{xi}\} \overset{\infty}{\longleftrightarrow} \{u_{yi} \overset{\kappa_y}{\longleftrightarrow} u_{yj}\} \overset{\infty}{\longleftrightarrow} \{u_{zi} \overset{\kappa_z}{\longleftrightarrow} u_{zj}\}$<br>Heterogeneous correlations.<br>Alike dof: ($\kappa_x$,$\kappa_y$,$\kappa_z$); Different dof: $\kappa\to\infty$. | $\pi^{-\frac{3}{2}}\cdot\frac{(\kappa_x-\frac{3}{2})^{-\frac{1}{2}}\Gamma(\kappa_x)}{\Gamma(\kappa_x-\frac{1}{2})}\cdot\frac{(\kappa_y-\frac{3}{2})^{-\frac{1}{2}}\Gamma(\kappa_y)}{\Gamma(\kappa_y-\frac{1}{2})}\cdot\frac{(\kappa_z-\frac{3}{2})^{-\frac{1}{2}}\Gamma(\kappa_z)}{\Gamma(\kappa_z-\frac{1}{2})}\cdot\theta_x^{-1}\theta_y^{-1}\theta_z^{-1}$<br>$\times\left(1+\frac{1}{\kappa_x-\frac{3}{2}}\cdot\frac{u_x^{2}}{\theta_x^{2}}\right)^{-\kappa_x}\cdot\left(1+\frac{1}{\kappa_y-\frac{3}{2}}\cdot\frac{u_y^{2}}{\theta_y^{2}}\right)^{-\kappa_y}\cdot\left(1+\frac{1}{\kappa_z-\frac{3}{2}}\cdot\frac{u_z^{2}}{\theta_z^{2}}\right)^{-\kappa_z}$ |
| 3 | $\left\{\{u_{\perp xi} \overset{\kappa}{\longleftrightarrow} u_{\perp xj}\} \overset{\kappa}{\longleftrightarrow} \{u_{\perp yi} \overset{\kappa}{\longleftrightarrow} u_{\perp yj}\}\right\} \overset{\infty}{\longleftrightarrow} \{u_{\parallel i} \overset{\kappa}{\longleftrightarrow} u_{\parallel j}\}$<br>Heterogeneous correlations.<br>Alike dof: $\kappa$; Different dof: $\kappa$ between perpendicular *x* and *y*,<br>$\kappa\to\infty$ between perpendicular and parallel. | $\pi^{-\frac{3}{2}}(\kappa-\tfrac{3}{2})^{-\frac{3}{2}}\cdot\frac{\Gamma(\kappa)(\kappa-\frac{1}{2})}{\Gamma(\kappa-\frac{1}{2})}\cdot\theta_{\parallel}^{-1}\theta_{\perp}^{-2}\left(1+\frac{1}{\kappa-\frac{3}{2}}\cdot\frac{u_{\perp}^{2}}{\theta_{\perp}^{2}}\right)^{-\kappa-\frac{1}{2}}\cdot\left(1+\frac{1}{\kappa-\frac{3}{2}}\cdot\frac{u_{\parallel}^{2}}{\theta_{\parallel}^{2}}\right)^{-\kappa}$ |
| 4a | $\left\{\{u_{\perp xi} \overset{\kappa}{\longleftrightarrow} u_{\perp xj}\} \overset{\kappa}{\longleftrightarrow} \{u_{\perp yi} \overset{\kappa}{\longleftrightarrow} u_{\perp yj}\}\right\} \overset{\kappa^{\text{int}}}{\longleftrightarrow} \{u_{\parallel i} \overset{\kappa}{\longleftrightarrow} u_{\parallel j}\}$<br>Heterogeneous correlations.<br>Alike dof: $\kappa$; Different dof: $\kappa$ between perpendicular *x* and *y*,<br>$\kappa^{\text{int}}$ between perpendicular and parallel. | $\mathrm{C}\cdot\theta_{\perp}^{-2}\theta_{\parallel}^{-1}\cdot\left\{-1+\left(1+\frac{1}{\kappa-\frac{3}{2}}\cdot\frac{u_{\perp}^{2}}{\theta_{\perp}^{2}}\right)^{\frac{\kappa+\frac{1}{2}}{\kappa^{\text{int}}+\frac{1}{2}}}+\left(1+\frac{1}{\kappa-\frac{3}{2}}\cdot\frac{u_{\parallel}^{2}}{\theta_{\parallel}^{2}}\right)^{\frac{\kappa}{\kappa^{\text{int}}}}\right\}^{-\kappa^{\text{int}}-1}$ (**) |
| 4b | $\left\{\{u_{\perp xi} \overset{\kappa_{\perp x}}{\longleftrightarrow} u_{\perp xj}\} \overset{\kappa_{\perp}^{\text{int}}}{\longleftrightarrow} \{u_{\perp yi} \overset{\kappa_{\perp y}}{\longleftrightarrow} u_{\perp yj}\}\right\} \overset{\kappa_{\perp\parallel}^{\text{int}}}{\longleftrightarrow} \{u_{\parallel i} \overset{\kappa_{\parallel}}{\longleftrightarrow} u_{\parallel j}\}$<br>Heterogeneous correlations.<br>Alike dof: $(\kappa_{\perp x},\kappa_{\perp y},\kappa_{\parallel})$; Different dof: $\kappa_{\perp}^{\text{int}}$ between perpendicular *x*<br>and *y*, $\kappa_{\perp\parallel}^{\text{int}}$ between perpendicular and parallel. | $\mathrm{C}\cdot\theta_{\perp}^{-2}\theta_{\parallel}^{-1}\cdot\left\{-1+\left[1+\frac{u_{\parallel}^{2}}{(\kappa_{\parallel}-\frac{3}{2})\theta_{\parallel}^{2}}\right]^{\frac{\kappa_{\parallel}}{\kappa_{\perp\parallel}^{\text{int}}}}+\left\{-1+\left[1+\frac{u_{\perp x}^{2}}{(\kappa_{\perp x}-\frac{3}{2})\theta_{\perp x}^{2}}\right]^{\frac{\kappa_{\perp x}}{\kappa_{\perp}^{\text{int}}}}+\left[1+\frac{u_{\perp y}^{2}}{(\kappa_{\perp y}-\frac{3}{2})\theta_{\perp y}^{2}}\right]^{\frac{\kappa_{\perp y}}{\kappa_{\perp}^{\text{int}}}}\right\}^{\frac{\kappa_{\perp}^{\text{int}}+\frac{1}{2}}{\kappa_{\perp\parallel}^{\text{int}}+\frac{1}{2}}}\right\}^{-\kappa_{\perp\parallel}^{\text{int}}-1}$ |

**Notes.** * (1) Homogeneous correlations. (2) Heterogeneous correlations, among alike dof: (a) $\kappa$, (b) ($\kappa_x$,$\kappa_y$,$\kappa_z$); among different dof: $\kappa\to\infty$. (3) Heterogeneous correlations, among alike dof: $\kappa$; among different dof: $\kappa$ between perpendicular *x* and *y*, $\kappa\to\infty$ between perpendicular and parallel. (4) Heterogeneous correlations, among alike dof: (a) $\kappa$, (b) $(\kappa_{\perp x},\kappa_{\perp y},\kappa_{\parallel})$; among different dof: (a) $\kappa$ between perpendicular *x* and *y*, $\kappa^{\text{int}}$ between perpendicular and parallel, (b) $\kappa_{\perp}^{\text{int}}$ between perpendicular *x* and *y*, $\kappa_{\perp\parallel}^{\text{int}}$ between perpendicular and parallel. ** The normalization constant C of the last two examples does not exist in a closed-form expression; it can be given in a series expression and derived numerically (see Appendix A).

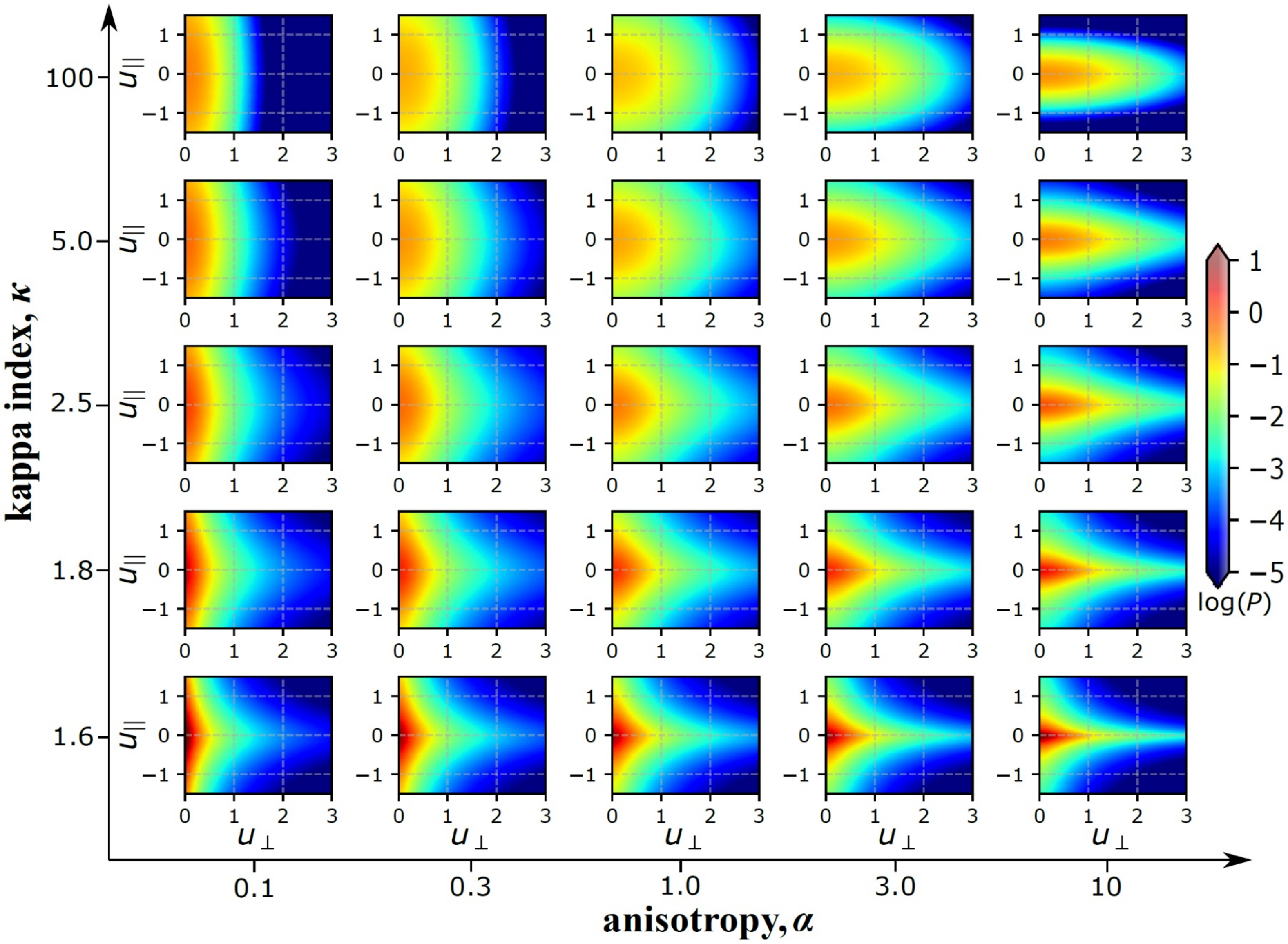

**Figure 4.** Anisotropic kappa distributions with heterogeneous correlation equal to zero, $\kappa^{\text{int}}\to\infty$, for various kappa indices $\kappa$ and anisotropies $\alpha$.

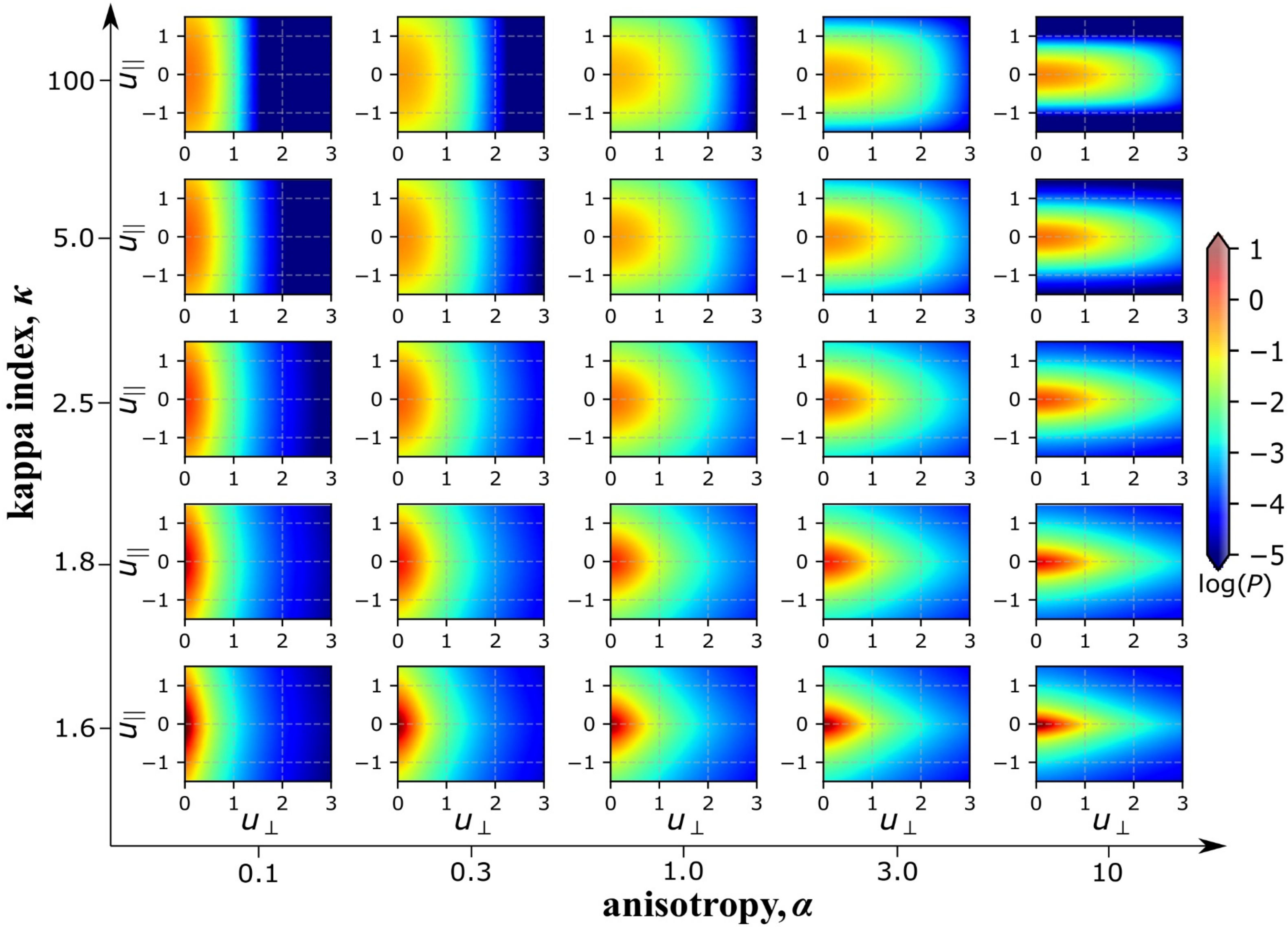

**Figure 5.** Anisotropic kappa distributions with heterogeneous correlation corresponding to $\kappa^{int}$=3, for various kappa indices $\kappa$ and anisotropies $\alpha$.

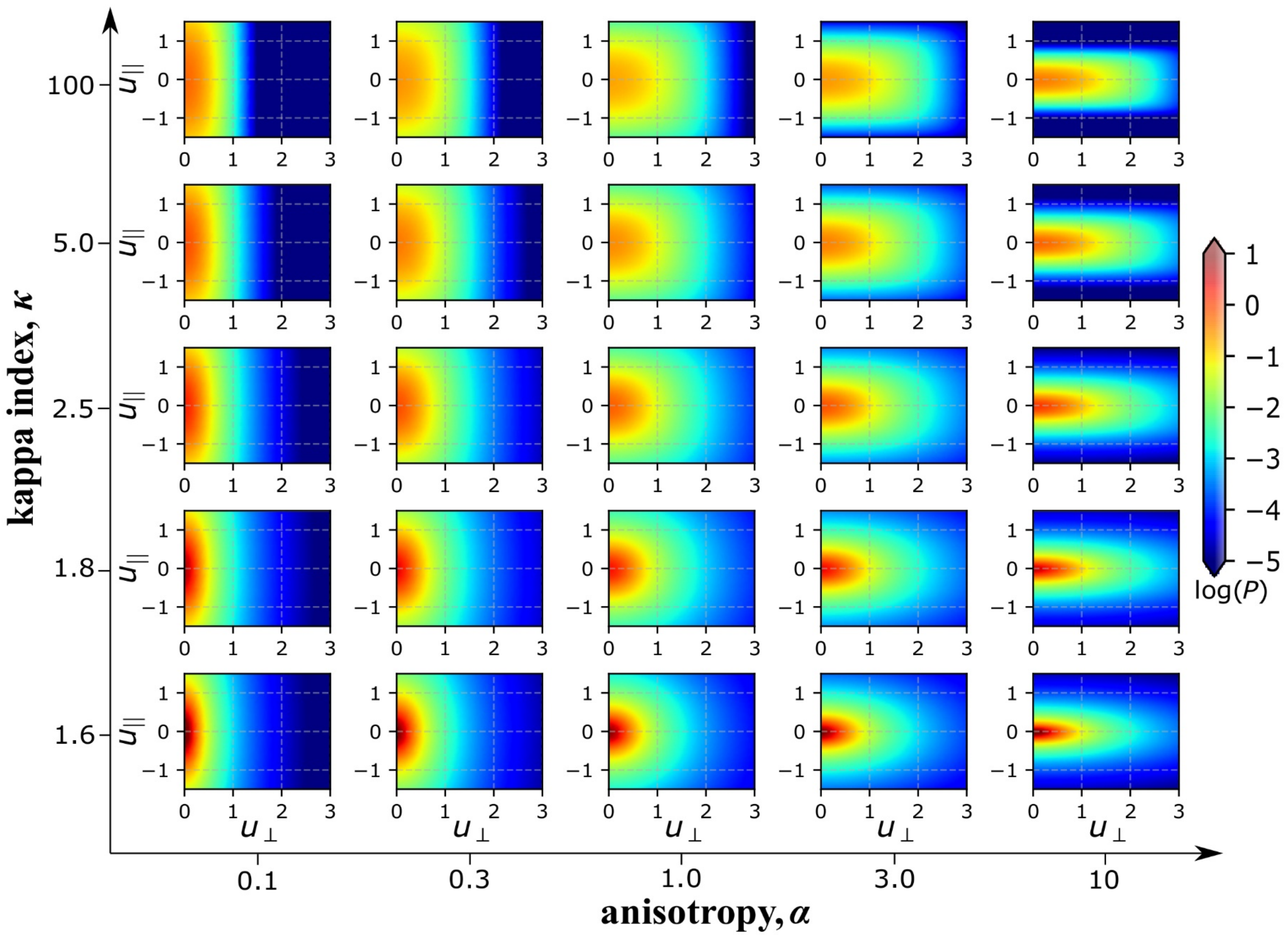

**Figure 6.** Anisotropic kappa distributions with heterogeneous correlation corresponding to $\kappa^{\mathrm{int}}$=1.5, for various kappa indices $\kappa$ and anisotropies $\alpha$.

## 6. Formalism of anisotropic kappa distributions in arbitrary reference frames

*6.1. Distributions of energy E, pitch angle ϑ, and azimuth φ*

We derive the expressions of the anisotropic kappa distributions in an arbitrary reference frame. We focus on the three main distributions shown in Section §5.4; there, the distribution had been expressed in the comoving system (where the bulk velocity is set to zero); using the transformations $u_{\parallel} = u\cos\vartheta$, $u_{\perp x} = u\sin\vartheta\cos\varphi$, $u_{\perp y} = u\sin\vartheta\sin\varphi$, (i.e., $\cos\vartheta = u_{\parallel}/u$ and $\cos\varphi = u_{\perp x}/u_{\perp}$), with $E = \frac{1}{2}mu^2$ (kinetic energy in the S/C reference frame), and $u_{\mathrm{b}\parallel} = u_{\mathrm{b}}\cos\vartheta_{\mathrm{b}}$, $u_{\mathrm{b}\perp x} = u_{\mathrm{b}}\sin\vartheta_{\mathrm{b}}\cos\varphi_{\mathrm{b}}$, $u_{\mathrm{b}\perp y} = u_{\mathrm{b}}\sin\vartheta_{\mathrm{b}}\sin\varphi_{\mathrm{b}}$ with $E_{\mathrm{b}} = \frac{1}{2}mu_{\mathrm{b}}^2$ (bulk kinetic energy), we express the distributions as follows:

- Homogeneous correlations (correlations between velocity components of each particle, are equal to the correlations between the same components of different particles, $\kappa^{\mathrm{int}}=\kappa$):

$$P(\vec{u}-\vec{u}_{\mathrm{b}};\theta,\alpha,\kappa;\kappa^{\mathrm{int}}=\kappa) = \pi^{-\frac{3}{2}}(\kappa-\tfrac{3}{2})^{-\frac{3}{2}}\frac{\Gamma(\kappa+1)}{\Gamma(\kappa-\frac{1}{2})}\cdot\alpha^{-1}[\tfrac{1}{3}(1+2\alpha)]^{\frac{3}{2}}\theta^{-3}$$
$$\times\left\{1+\frac{1}{\kappa-\frac{3}{2}}\cdot\frac{1+2\alpha}{3\alpha\theta^2}\cdot\left[\alpha(u_{\parallel}-u_{\mathrm{b}\parallel})^2+(u_{\perp}-u_{\mathrm{b}\perp})^2\right]\right\}^{-\kappa-1}, \tag{71a}$$

or, in terms of the kinetic energy $E$, pitch angle $\vartheta$, and azimuth angle $\varphi$:

$$P(E,\vartheta,\varphi;E_{\mathrm{b}},\vartheta_{\mathrm{b}},\varphi_{\mathrm{b}};T,\alpha,\kappa;\kappa^{\mathrm{int}}=\kappa) = \pi^{-\frac{3}{2}}(\kappa-\tfrac{3}{2})^{-\frac{3}{2}}\frac{\Gamma(\kappa+1)}{\Gamma(\kappa-\frac{1}{2})}$$
$$\times\alpha^{-1}[\tfrac{1}{3}(1+2\alpha)]^{\frac{3}{2}}(2k_{\mathrm{B}}T/m)^{-\frac{3}{2}}\cdot\left[1+\frac{1}{\kappa-\frac{3}{2}}\cdot\frac{1+2\alpha}{3\alpha k_{\mathrm{B}}T}\cdot L(E,\vartheta,\varphi;E_{\mathrm{b}},\vartheta_{\mathrm{b}},\varphi_{\mathrm{b}};\alpha)\right]^{-\kappa-1}, \tag{71b}$$

where

$$L(E,\vartheta,\varphi;E_{\mathrm{b}},\vartheta_{\mathrm{b}},\varphi_{\mathrm{b}};\alpha) = E[1+(\alpha-1)\cos^2\vartheta]+E_{\mathrm{b}}[1+(\alpha-1)\cos^2\vartheta_{\mathrm{b}}]$$
$$-2\sqrt{E\,E_{\mathrm{b}}}\,[\alpha\cos\vartheta\cos\vartheta_{\mathrm{b}}+\sin\vartheta\sin\vartheta_{\mathrm{b}}\cos(\varphi-\varphi_{\mathrm{b}})]\ . \tag{71c}$$

- Heterogeneous correlations equal to zero, (no correlations between velocity components of each particle, $\kappa^{\mathrm{int}}\to\infty$, and finite correlations between the same components of different particles, $\kappa<\infty$):

$$P(\vec{u}-\vec{u}_{\mathrm{b}};\theta,\alpha,\kappa;\kappa^{\mathrm{int}}\to\infty) = [\pi(\kappa-\tfrac{3}{2})]^{-\frac{3}{2}}\cdot\frac{\Gamma(\kappa)(\kappa-\frac{1}{2})}{\Gamma(\kappa-\frac{1}{2})}\cdot\alpha^{-1}[\tfrac{1}{3}(1+2\alpha)]^{\frac{3}{2}}\theta^{-3}$$
$$\times\left[1+\frac{1}{\kappa-\frac{3}{2}}\cdot\frac{1+2\alpha}{3\alpha\theta^2}\cdot(u_{\perp}-u_{\mathrm{b}\perp})^2\right]^{-\kappa-\frac{1}{2}}\cdot\left[1+\frac{1}{\kappa-\frac{3}{2}}\cdot\frac{1+2\alpha}{3\theta^2}\cdot(u_{\parallel}-u_{\mathrm{b}\parallel})^2\right]^{-\kappa}, \tag{72a}$$

or, in terms of the kinetic energy $E$, pitch angle $\vartheta$, and azimuth angle $\varphi$:

$$P(E,\vartheta,\varphi;E_b,\vartheta_b,\varphi_b;T,\alpha,\kappa;\kappa^{int}\to\infty)=[\pi(\kappa-\tfrac{3}{2})]^{-\frac{3}{2}}\cdot\frac{\Gamma(\kappa)(\kappa-\frac{1}{2})}{\Gamma(\kappa-\frac{1}{2})}\cdot\alpha^{-1}[\tfrac{1}{3}(1+2\alpha)]^{\frac{3}{2}}(2k_BT/m)^{-\frac{3}{2}}$$

$$\times\left\{1+\frac{1}{\kappa-\frac{3}{2}}\cdot\frac{1+2\alpha}{3\alpha k_BT}\cdot\left[E\sin^2\vartheta+E_b\sin^2\vartheta_b-2\sqrt{E\,E_b}\sin\vartheta\sin\vartheta_b\cos(\varphi-\varphi_b)\right]\right\}^{-\kappa-\frac{1}{2}} \quad (72b)$$

$$\times\left[1+\frac{1}{\kappa-\frac{3}{2}}\cdot\frac{1+2\alpha}{3k_BT}\cdot(E\cos^2\vartheta+E_b\cos^2\vartheta_b-2\sqrt{E\,E_b}\cos\vartheta\cos\vartheta_b)\right]^{-\kappa}.$$

- Arbitrary heterogeneous correlation (finite correlations between velocity components of each particle, $\kappa^{int}<\infty$, and finite correlations between the same components of different particles, $\kappa<\infty$):

$$P(\vec{u}-\vec{u}_b;\theta,\alpha,\kappa;\kappa^{int})=\mathrm{A}(\kappa^{int},\kappa)\cdot\alpha^{-1}[\tfrac{1}{3}(1+2\alpha)]^{\frac{3}{2}}\theta^{-3}$$

$$\times\left\{\begin{array}{l}-1+\left[1+\dfrac{1}{\kappa-\frac{3}{2}}\cdot\dfrac{1+2\alpha}{3\theta^2}\cdot(u_{\parallel}-u_{b\parallel})^2\right]^{\frac{\kappa}{\kappa^{int}}}\\ +\left[1+\dfrac{1}{\kappa-\frac{3}{2}}\cdot\dfrac{1+2\alpha}{3\alpha\theta^2}\cdot(u_{\perp}-u_{b\perp})^2\right]^{\frac{\kappa+\frac{1}{2}}{\kappa^{int}+\frac{1}{2}}}\end{array}\right\}^{-\kappa^{int}-1}, \quad (73a)$$

or, in terms of the kinetic energy $E$, pitch angle $\vartheta$, and azimuth angle $\varphi$:

$$P(E,\vartheta,\varphi;E_b,\vartheta_b,\varphi_b;T,\alpha,\kappa;\kappa^{int})=\mathrm{A}(\kappa^{int},\kappa)\cdot\alpha^{-1}[\tfrac{1}{3}(1+2\alpha)]^{\frac{3}{2}}(2k_BT/m)^{-\frac{3}{2}}$$

$$\times\left\{-1+\left[1+\frac{1}{\kappa-\frac{3}{2}}\cdot\frac{1+2\alpha}{3k_BT}\cdot\left(E\cos^2\vartheta+E_b\cos^2\vartheta_b-2\sqrt{E\,E_b}\cos\vartheta\cos\vartheta_b\right)\right]^{\frac{\kappa}{\kappa^{int}}}\right. \quad (73b)$$

$$\left.+\left\{1+\frac{1}{\kappa-\frac{3}{2}}\cdot\frac{1+2\alpha}{3\alpha k_BT}\cdot\left[E\sin^2\vartheta+E_b\sin^2\vartheta_b-2\sqrt{E\,E_b}\sin\vartheta\sin\vartheta_b\cos(\varphi-\varphi_b)\right]\right\}^{\frac{\kappa+\frac{1}{2}}{\kappa^{int}+\frac{1}{2}}}\right\}^{-\kappa^{int}-1}.$$

Note that in all the above equations, we used $(u_{\perp}-u_{b\perp})^2=(u_{\perp x}-u_{b\perp x})^2+(u_{\perp y}-u_{b\perp y})^2$.

Figure 7 plots the distribution (71b), corresponding to the special case of homogeneous correlations among velocity components of correlated particles, i.e., for $\kappa^{int}=\kappa$; Figure 8 plots the distribution (72b), i.e., corresponding to the special case of of zero heterogeneous correlations among velocity components, i.e., $\kappa^{int}\to\infty$; Figure 9 plots the distribution (73b), that is, one of the nontrivial cases of finite heterogeneous correlations, i.e., $\kappa^{int}\neq\kappa$.

We observe various peaks of the azimuth independent distribution appear at several pitch angles, which become clearer for anisotropies closer to 1. This behavior opposes the smoother plots characterizing the respective distributions in their comoving reference frame (Figures 4-6). The explanation is that, in each of the panels of Figures 7-9, the azimuth angle is fixed to a certain value.

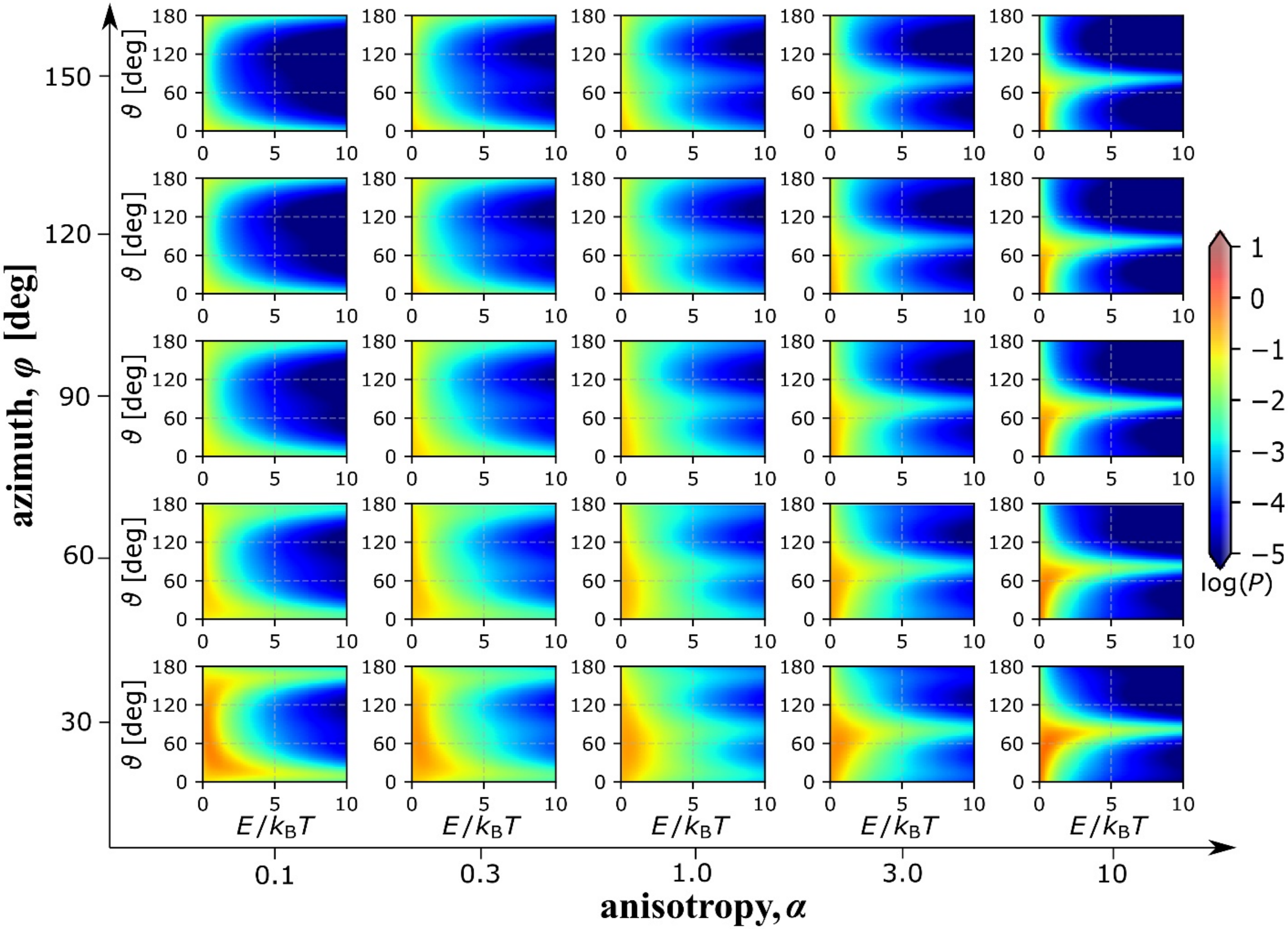

**Figure 7.** Anisotropic kappa distributions with heterogeneous correlation equal to zero, $\kappa^{int}\rightarrow\infty$, for $\vartheta_b$=60$^0$, $E_b/(k_BT)$=0.5, kappa index $\kappa$=3 and various anisotropies $\alpha$, in an arbitrary reference frame.

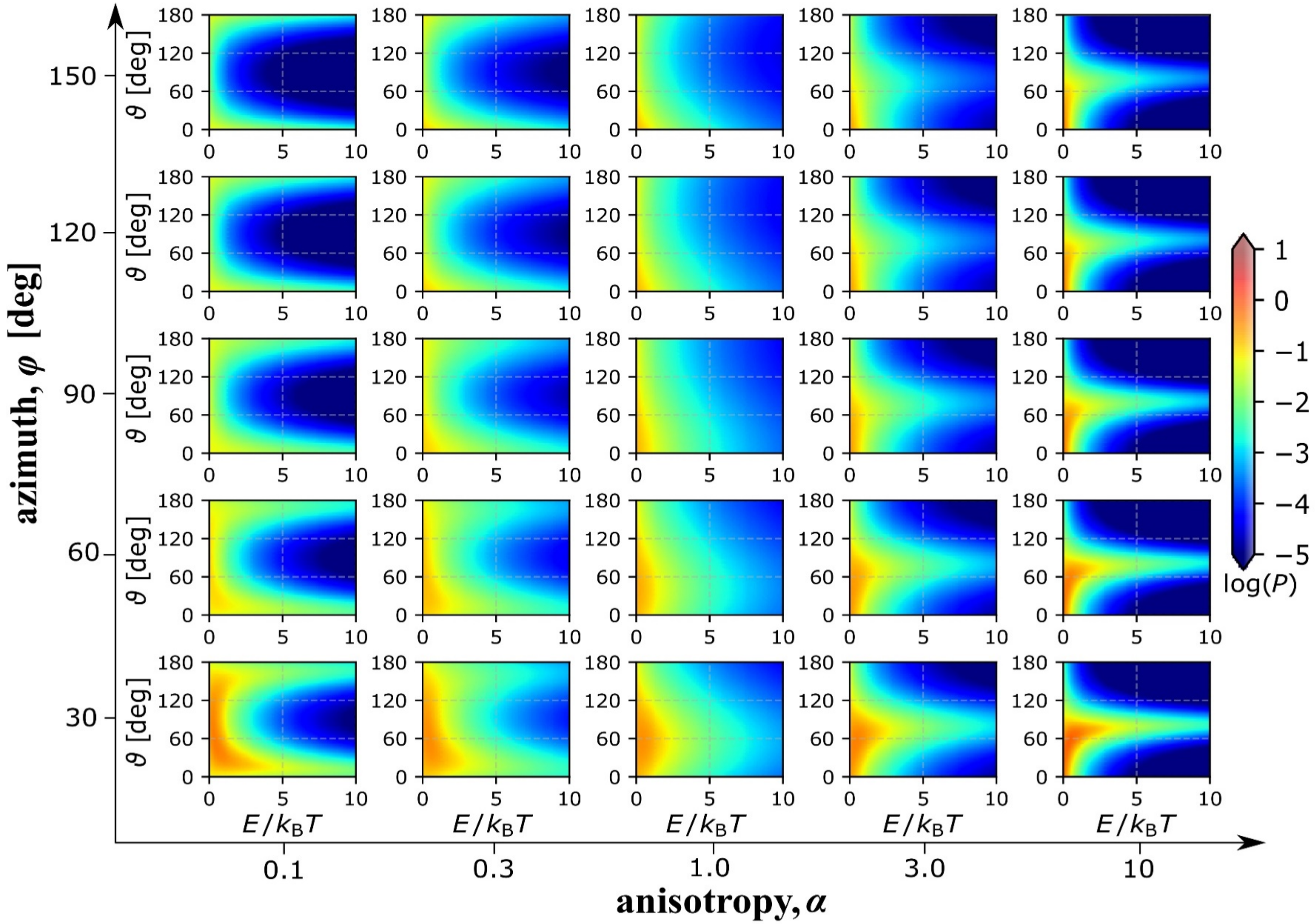

**Figure 8.** Anisotropic kappa distributions with heterogeneous correlation corresponding to $\kappa^{\text{int}}$=3, for $\vartheta_b$=60$^0$, $E_b/(k_B T)$=0.5, kappa index $\kappa$=3 and various anisotropies $\alpha$, in an arbitrary reference frame.

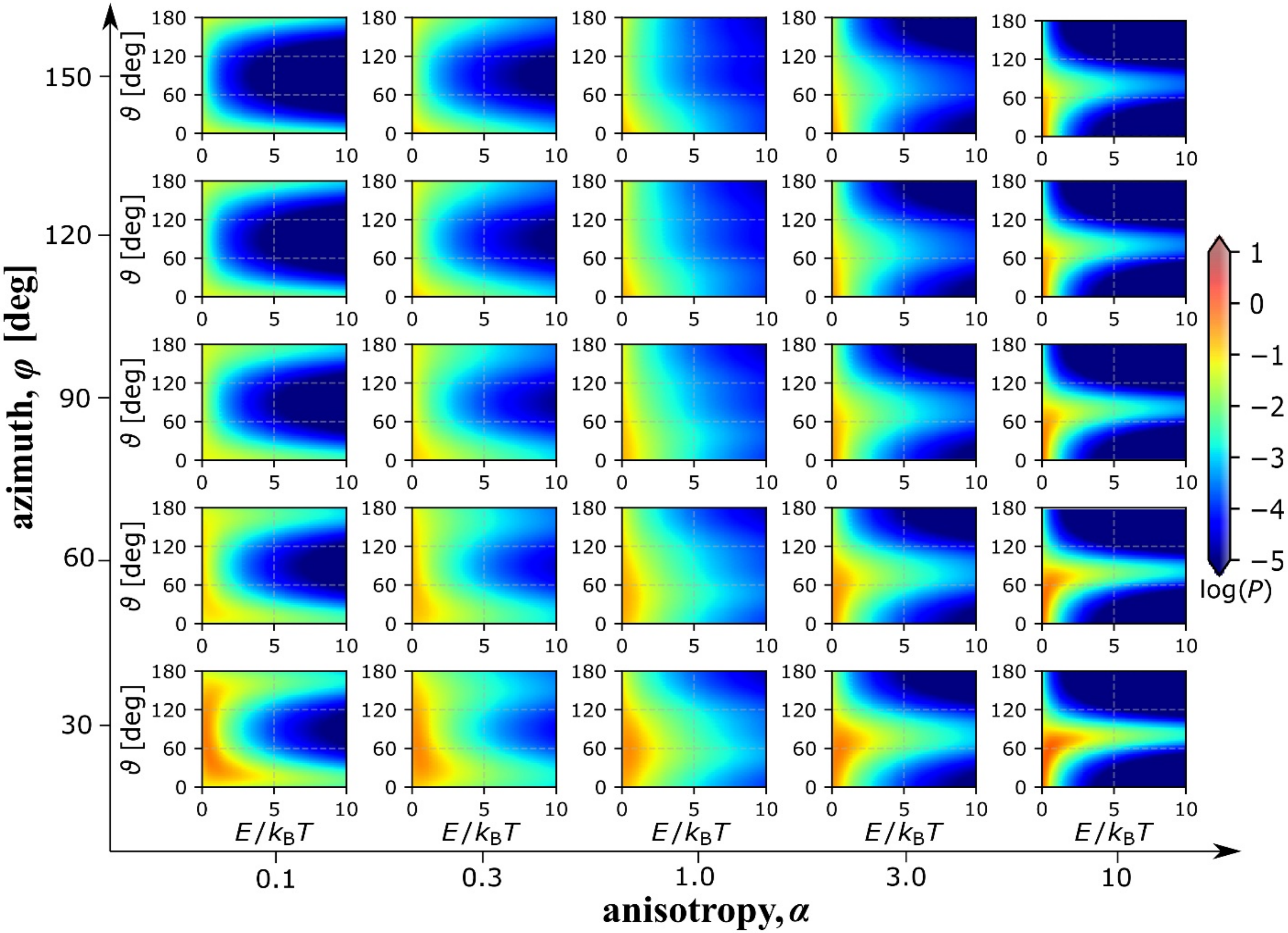

**Figure 9.** Anisotropic kappa distributions with heterogeneous correlation corresponding to $\kappa^{\mathrm{int}}$=1.5, for $\vartheta_{\mathrm{b}}$=60$^0$, $E_{\mathrm{b}}/(k_{\mathrm{B}}T)$=0.5, kappa index $\kappa$=3 and various anisotropies $\alpha$, in an arbitrary reference frame.

*6.2. Azimuth-integrated distributions of energy E and pitch angle $\vartheta$*

Measurements taken from instruments that collect at certain energy and pitch angle channels can be used to fit the theoretical distributions, however, the latter must be azimuth-independent. For this, we use the azimuth-integrated distributions, that is, functions of energy $E$ and pitch angle $\vartheta$, and not of azimuth $\varphi$. These are derived as follows:

In general, for the azimuthal function $f(\varphi)$, we find its expectation value, i.e., its average, as : $<f> = (1/2\pi) \int f(\varphi)\, d\varphi$, that is,

$$P(E,\vartheta) = \tfrac{1}{2\pi}\int_0^{2\pi} P(E,\vartheta,\varphi)d\varphi \ . \tag{74a}$$

Moreover, since we deal with an expectation value, it is reasonable to also seek for the standard deviation, or the 1-sigma uncertainty, given by $\delta<f> = \surd\{(1/2\pi) \int [f(\varphi)- <f>]^2\, d\varphi\}$, that is

$$\delta P(E,\vartheta) = \sqrt{\tfrac{1}{2\pi}\int_0^{2\pi}[P(E,\vartheta,\varphi) - P(E,\vartheta)]^2 d\varphi} \ . \tag{74b}$$

Below we apply Eqs.(74a,b) to each of the $(E,\vartheta,\varphi)$-depended distributions expressed by Eqs.(71b,72b,73b). (Note that we neglected the dependence on parameters, $E_b$, $\vartheta_b$, $T$, $\alpha$, $\kappa$, for simplicity).

- Homogeneous correlations, $\kappa^{int}=\kappa$:

The azimuth-independent kappa distribution of energy $E$ and pitch angle $\vartheta$, corresponding to Eq.(71b), is given by:

$$\begin{aligned} P(E,\vartheta;\kappa^{int}\to\kappa) &= [\pi(\kappa-\tfrac{3}{2})(\tfrac{2k_B T}{m})]^{-\frac{3}{2}} \frac{\Gamma(\kappa+1)}{\Gamma(\kappa-\frac{1}{2})}\alpha^{\frac{1}{2}}(\tfrac{1+2\alpha}{3a})^{\frac{1}{2}-\kappa} \\ &\times\left[\frac{\sqrt{EE_b}\sin\vartheta\sin\vartheta_b}{(\kappa-\frac{3}{2})k_B T\cdot \mathrm{A}(\lambda)}\right]^{-\kappa-1} \cdot {}_2F_1\left[\kappa+1,\kappa+1,1,\mathrm{A}(\lambda)^2\right] \ , \end{aligned} \tag{75a}$$

where we have set $\mathrm{A} = \mathrm{A}[\lambda(E,\vartheta)]$ as:

$$\mathrm{A}(x) \equiv x - \sqrt{x^2-1} \ \text{, and} \tag{75b}$$

$$\lambda(E,\vartheta) \equiv -\alpha\tan\vartheta\tan\vartheta_b + \frac{(\kappa-\frac{3}{2})k_B T\,\frac{3\alpha}{2\alpha+1} + E[(\alpha-1)\cos^2\vartheta+1] + E_b[(\alpha-1)\cos^2\vartheta_b+1]}{2\sqrt{EE_b}\sin\vartheta\sin\vartheta_b} \ . \tag{75c}$$

Then, the theoretical uncertainty, caused by substituting the azimuth with its expectation value, is

$$\begin{aligned} \delta P(E,\vartheta;\kappa^{int}\to\kappa) &= [\pi(\kappa-\tfrac{3}{2})(\tfrac{2k_B T}{m})]^{-\frac{3}{2}} \frac{\Gamma(\kappa+1)}{\Gamma(\kappa-\frac{1}{2})}\alpha^{\frac{1}{2}}(\tfrac{1+2\alpha}{3a})^{\frac{1}{2}-\kappa} \\ &\times\left[\frac{\sqrt{EE_b}\sin\vartheta\sin\vartheta_b}{(\kappa-\frac{3}{2})k_B T\cdot \mathrm{A}(\lambda)}\right]^{-\kappa-1} \\ &\times\sqrt{{}_2F_1\left[2(\kappa+1),2(\kappa+1),1,\mathrm{A}(\lambda)^2\right] - {}_2F_1\left[\kappa+1,\kappa+1,1,\mathrm{A}(\lambda)^2\right]^2} \ . \end{aligned} \tag{75d}$$

- Heterogeneous correlations equal to zero, $\kappa^{\text{int}}\to\infty$:

The azimuth-independent kappa distribution of energy $E$ and pitch angle $\vartheta$, corresponding to Eq.(72b), is given by

$$P(E,\vartheta;\kappa^{\text{int}}\to\infty)=[\pi\,(\kappa-\tfrac{3}{2})(\tfrac{2k_{\text{B}}T}{m})]^{-\frac{3}{2}}\,\frac{\Gamma(\kappa)(\kappa-\frac{1}{2})}{\Gamma(\kappa-\frac{1}{2})}\,\alpha^{\frac{1}{2}}(\tfrac{1+2\alpha}{3a})^{1-\kappa}$$
$$\times\left[1+\frac{1}{\kappa-\frac{3}{2}}\cdot\frac{1+2\alpha}{3k_{\text{B}}T}\cdot\left(\sqrt{E}\cos\vartheta-\sqrt{E_{\text{b}}}\cos\vartheta_{\text{b}}\right)^{2}\right]^{-\kappa}\tag{76a}$$
$$\times\left[\frac{\sqrt{EE_{\text{b}}}\sin\vartheta\sin\vartheta_{\text{b}}}{(\kappa-\frac{3}{2})k_{\text{B}}T\cdot\mathrm{A}(\lambda)}\right]^{-\kappa-\frac{1}{2}}\cdot{}_{2}F_{1}\left[\kappa+\tfrac{1}{2},\kappa+\tfrac{1}{2},1,\mathrm{A}(\lambda)^{2}\right]\,.$$

The corresponding theoretical uncertainty, caused by substituting the azimuth with its expectation value, is now given by

$$\delta P(E,\vartheta;\kappa^{\text{int}}\to\infty)=[\pi\,(\kappa-\tfrac{3}{2})(\tfrac{2k_{\text{B}}T}{m})]^{-\frac{3}{2}}\,\frac{\Gamma(\kappa)(\kappa-\frac{1}{2})}{\Gamma(\kappa-\frac{1}{2})}\,\alpha^{\frac{1}{2}}(\tfrac{1+2\alpha}{3a})^{1-\kappa}$$
$$\times\left[1+\frac{1}{\kappa-\frac{3}{2}}\cdot\frac{1+2\alpha}{3k_{\text{B}}T}\cdot\left(\sqrt{E}\cos\vartheta-\sqrt{E_{\text{b}}}\cos\vartheta_{\text{b}}\right)^{2}\right]^{-\kappa}$$
$$\times\left[\frac{\sqrt{EE_{\text{b}}}\sin\vartheta\sin\vartheta_{\text{b}}}{(\kappa-\frac{3}{2})k_{\text{B}}T\cdot\mathrm{A}(\lambda)}\right]^{-\kappa-\frac{1}{2}}\tag{76b}$$
$$\times\sqrt{{}_{2}F_{1}\left[2\kappa+1,2\kappa+1,1,\mathrm{A}(\lambda)^{2}\right]-{}_{2}F_{1}\left[\kappa+\tfrac{1}{2},\kappa+\tfrac{1}{2},1,\mathrm{A}(\lambda)^{2}\right]^{2}}\quad,$$

where we have now set $\mathrm{A}=\mathrm{A}[\lambda(E,\vartheta)]$ as:

$$\mathrm{A}(x)\equiv x-\sqrt{x^{2}-1}\;,\tag{76c}$$

$$\lambda(E,\vartheta)\equiv\frac{(\kappa-\frac{3}{2})k_{\text{B}}T\,\frac{3\alpha}{2\alpha+1}+E\sin^{2}\vartheta+E_{\text{b}}\sin^{2}\vartheta_{\text{b}}}{2\sqrt{EE_{\text{b}}}\sin\vartheta\sin\vartheta_{\text{b}}}\;.\tag{76d}$$

- Arbitrary heterogeneous correlation, $\kappa^{\text{int}}<\infty$:

In the general case corresponding to Eq.(73b), the azimuth-independent kappa distribution of energy $E$ and pitch angle $\vartheta$ and the corresponding theoretical uncertainty are respectively given by:

$$P(E,\vartheta;\kappa^{\text{int}})=\mathrm{A}(\kappa^{\text{int}},\kappa)\cdot\alpha^{\frac{1}{2}}[\tfrac{3\alpha}{1+2\alpha}(\tfrac{2k_{\text{B}}T}{m})]^{-\frac{3}{2}}\cdot\pi^{-1}$$
$$\times\int_{0}^{\pi}\left\{\begin{array}{c}\left\{-1+\left[1+\frac{1}{\kappa-\frac{3}{2}}\cdot\frac{1+2\alpha}{3k_{\text{B}}T}\cdot\left(\sqrt{E}\cos\vartheta-\sqrt{E_{\text{b}}}\cos\vartheta_{\text{b}}\right)^{2}\right]^{\frac{\kappa}{\kappa^{\text{int}}}}\right.\\ \left.+\left\{1+\frac{1}{\kappa-\frac{3}{2}}\cdot\frac{1+2\alpha}{3\alpha k_{\text{B}}T}\cdot\left[E\sin^{2}\vartheta+E_{\text{b}}\sin^{2}\vartheta_{\text{b}}-2\sqrt{E\,E_{\text{b}}}\sin\vartheta\sin\vartheta_{\text{b}}\cos(X)\right]\right\}^{\frac{\kappa+\frac{1}{2}}{\kappa^{\text{int}}+\frac{1}{2}}}\right\}^{-\kappa^{\text{int}}-1}\end{array}\right\}dX\,,\tag{77a}$$

and

$$\delta P(E,\vartheta;\kappa^{\rm int}) = {\rm A}(\kappa^{\rm int},\kappa)\cdot\alpha^{\frac{1}{2}}[\tfrac{3\alpha}{1+2\alpha}(\tfrac{2k_{\rm B}T}{m})]^{-\frac{3}{2}}\cdot\pi^{-1}$$

$$\times\left\{\int_0^\pi\left\{\begin{array}{c}\left\{-1+\left[1+\frac{1}{\kappa-\frac{3}{2}}\cdot\frac{1+2\alpha}{3k_{\rm B}T}\cdot\left(\sqrt{E}\cos\vartheta-\sqrt{E_{\rm b}}\cos\vartheta_{\rm b}\right)^2\right]^{\frac{\kappa}{\kappa^{\rm int}}}\right.\\ \left.+\left\{1+\frac{1}{\kappa-\frac{3}{2}}\cdot\frac{1+2\alpha}{3\alpha k_{\rm B}T}\cdot\left[E\sin^2\vartheta+E_{\rm b}\sin^2\vartheta_{\rm b}-2\sqrt{E\,E_{\rm b}}\sin\vartheta\sin\vartheta_{\rm b}\cos(X)\right]\right\}^{\frac{\kappa+\frac{1}{2}}{\kappa^{\rm int}+\frac{1}{2}}}\right\}^{-\kappa^{\rm int}-1}\end{array}\right\}^2 dX \right.$$
$$\left.-\left\{\int_0^\pi\left\{\begin{array}{c}\left\{-1+\left[1+\frac{1}{\kappa-\frac{3}{2}}\cdot\frac{1+2\alpha}{3k_{\rm B}T}\cdot\left(\sqrt{E}\cos\vartheta-\sqrt{E_{\rm b}}\cos\vartheta_{\rm b}\right)^2\right]^{\frac{\kappa}{\kappa^{\rm int}}}\right.\\ \left.+\left\{1+\frac{1}{\kappa-\frac{3}{2}}\cdot\frac{1+2\alpha}{3\alpha k_{\rm B}T}\cdot\left[E\sin^2\vartheta+E_{\rm b}\sin^2\vartheta_{\rm b}-2\sqrt{E\,E_{\rm b}}\sin\vartheta\sin\vartheta_{\rm b}\cos(X)\right]\right\}^{\frac{\kappa+\frac{1}{2}}{\kappa^{\rm int}+\frac{1}{2}}}\right\}^{-\kappa^{\rm int}-1}\end{array}\right\}dX\right\}^2\right\}^{\frac{1}{2}} . \quad (77b)$$

Figures 10 and 11 plot the azimuth-independent kappa distribution of energy $E$ and pitch angle $\vartheta$, for both the cases of $\vartheta_b$=$0^0$ and $\vartheta_b$=$90^0$, respectively, for $\kappa^{int}$=$\kappa$, and for various values of the kappa index and anisotropy; (the case $\vartheta_b$=$180^0$ is symmetric to case of $\vartheta_b$=$0^0$ and is ignored). Similar are the plots in Figures 12 and 13, for $\kappa^{int}\rightarrow\infty$, and the plots in Figures 14 and 15, for for $\kappa^{int}$=1.5. We observe that the panels in Figures 10-15 are smooth compared to those in Figures 7-9, because the former figures are plotted after the integration of $\varphi$, while the latter figures are plotted for specific values of $\varphi$.

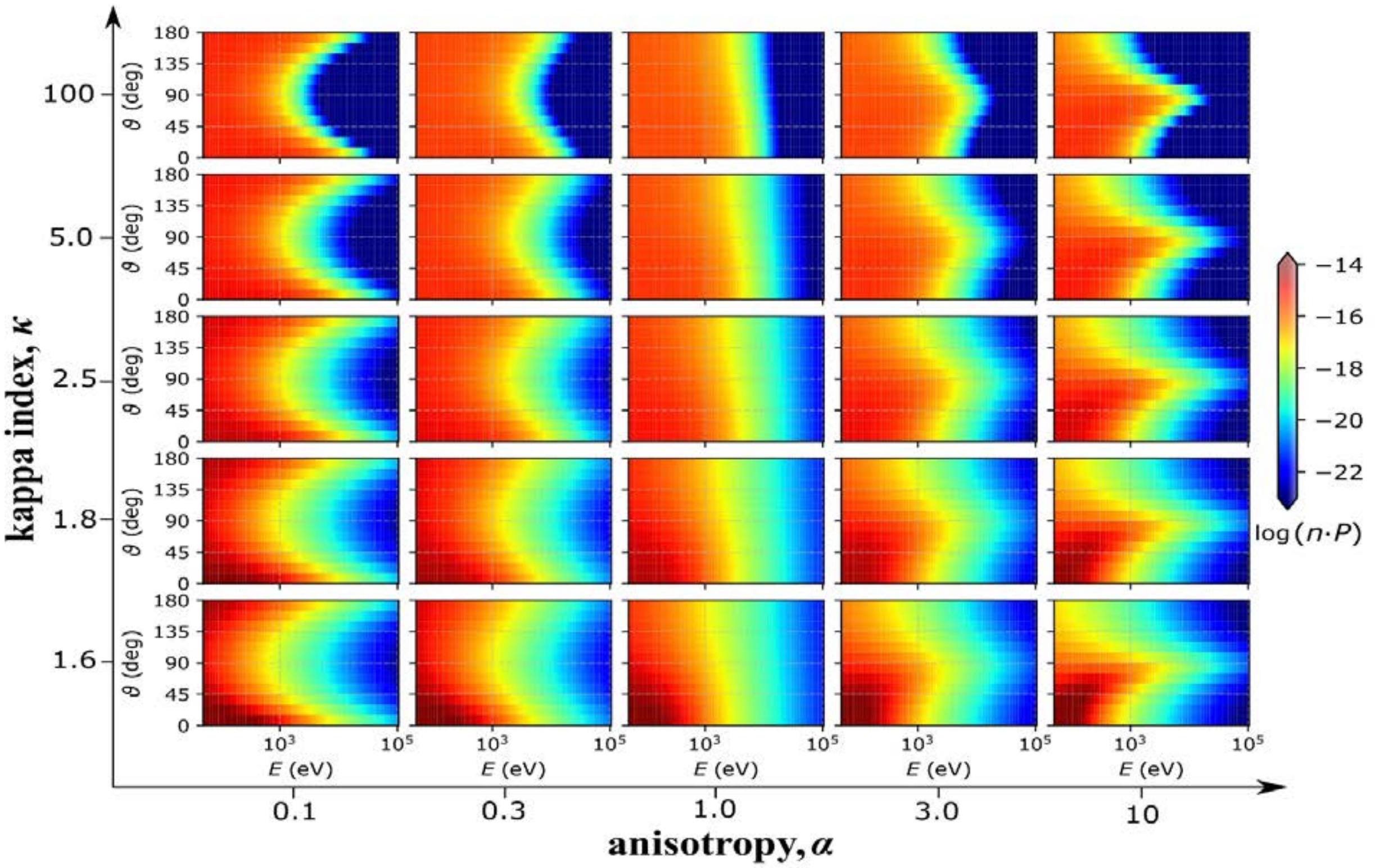

**Figure 10.** Azimuth-independent anisotropic kappa distribution of energy $E$ and pitch angle $\vartheta$, in an arbitrary reference frame, in the case of homogeneous correlations, Eq.(76a), i.e., $\kappa^{\rm int}=\kappa$, plotted for $\vartheta_{\rm b}=0^0$, $V_{\rm b}=5000$ km/s ($E_{\rm b}$ ~70 eV), $T=10^7$ K (~0.86 keV), and various anisotropies $\alpha$ and kappa indices $\kappa$.

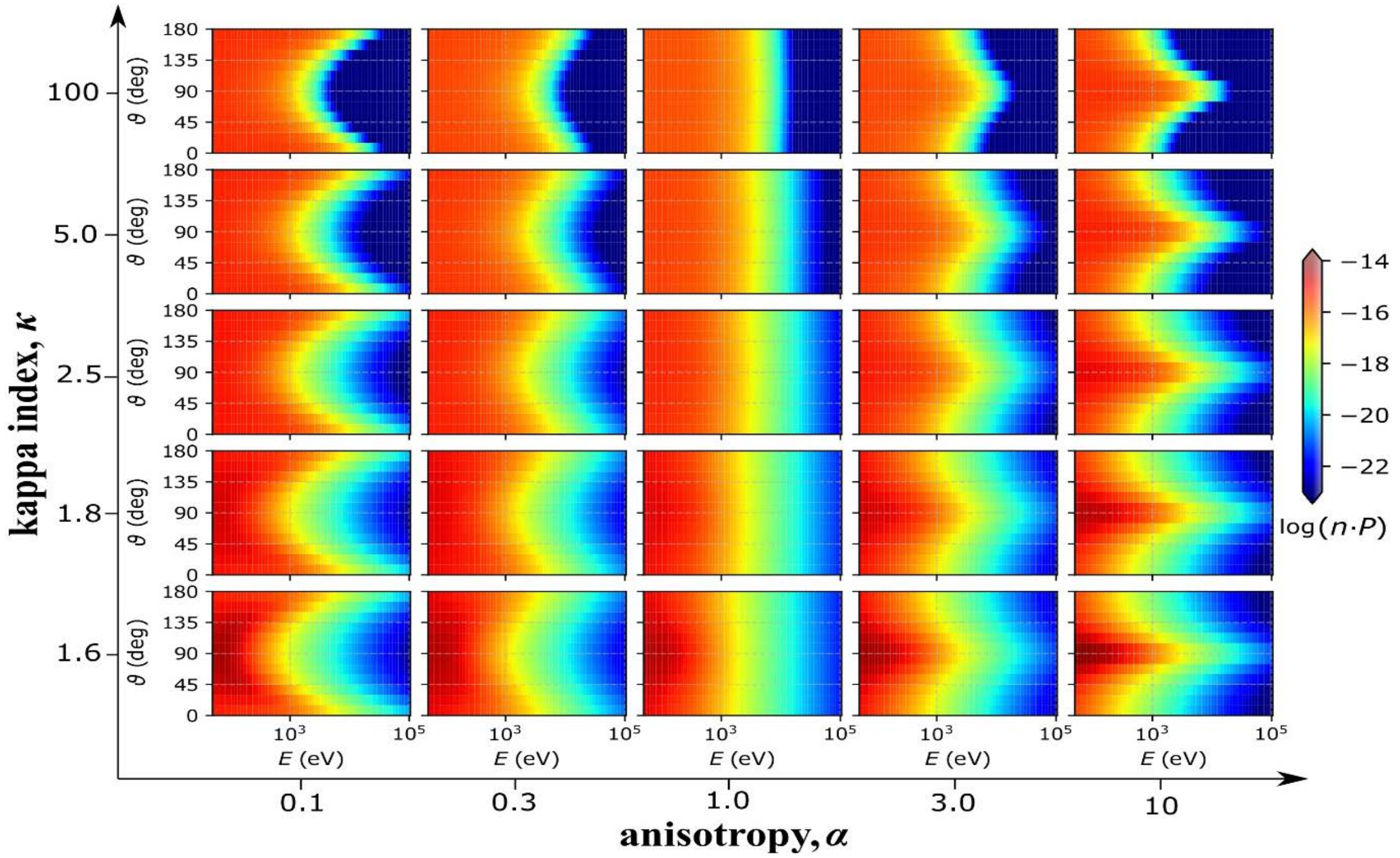

**Figure 11.** Similar to Figure 10 but for $\vartheta_{\rm b}=90^0$.

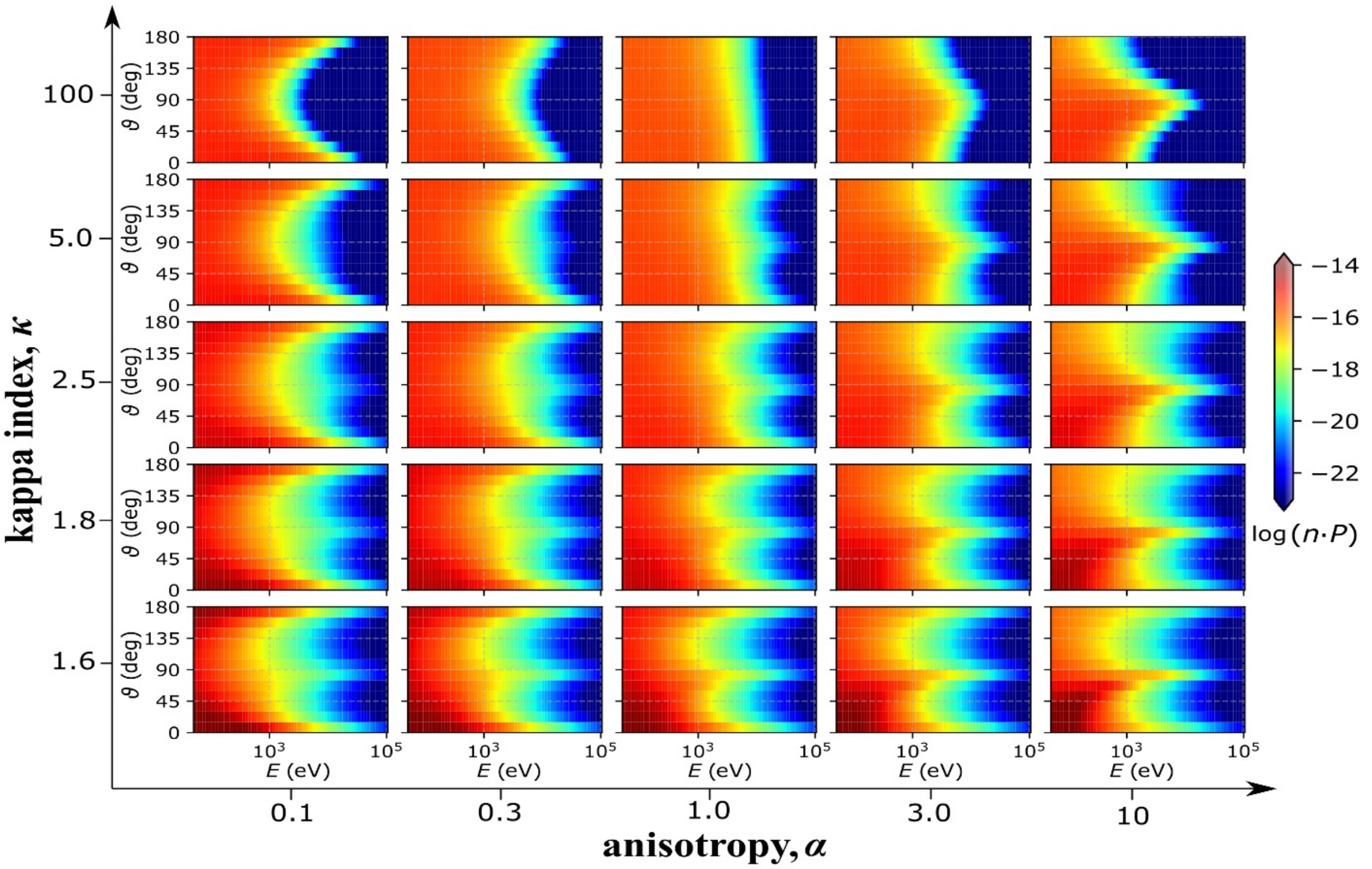

**Figure 12.** Azimuth-independent anisotropic kappa distribution of energy $E$ and pitch angle $\vartheta$, in an arbitrary reference frame, in the case of heterogeneous correlations, Eq.(77a), i.e., $\kappa^{\mathrm{int}} \to \infty$, plotted for $\vartheta_b = 0^0$, $V_b$=5000 km/s ($E_b$ ~70 eV), $T=10^7$ K (~0.86 keV), and various anisotropies $\alpha$ and kappa indices $\kappa$.

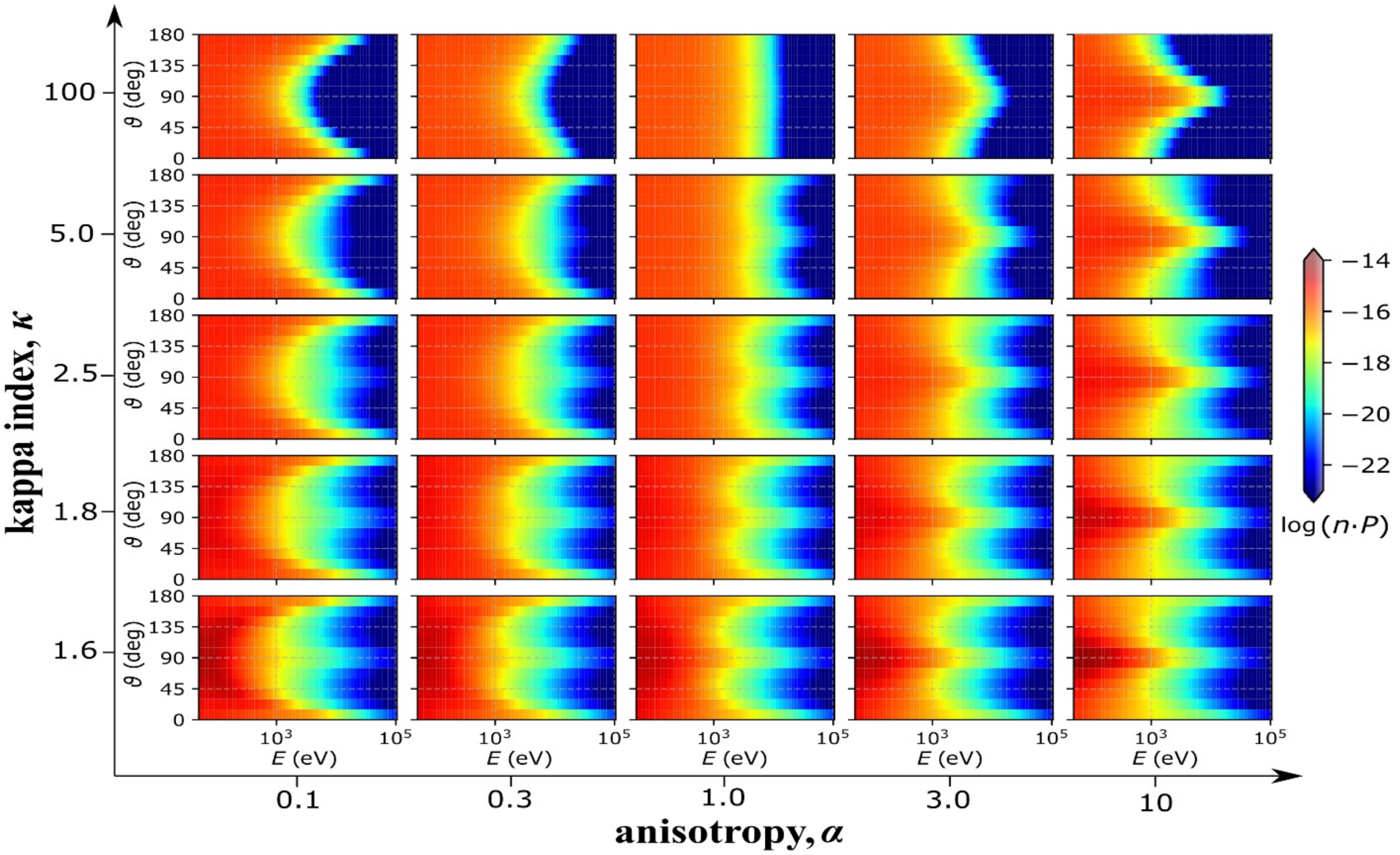

**Figure 13.** Similar to Figure 12 but for $\vartheta_b = 90^0$.

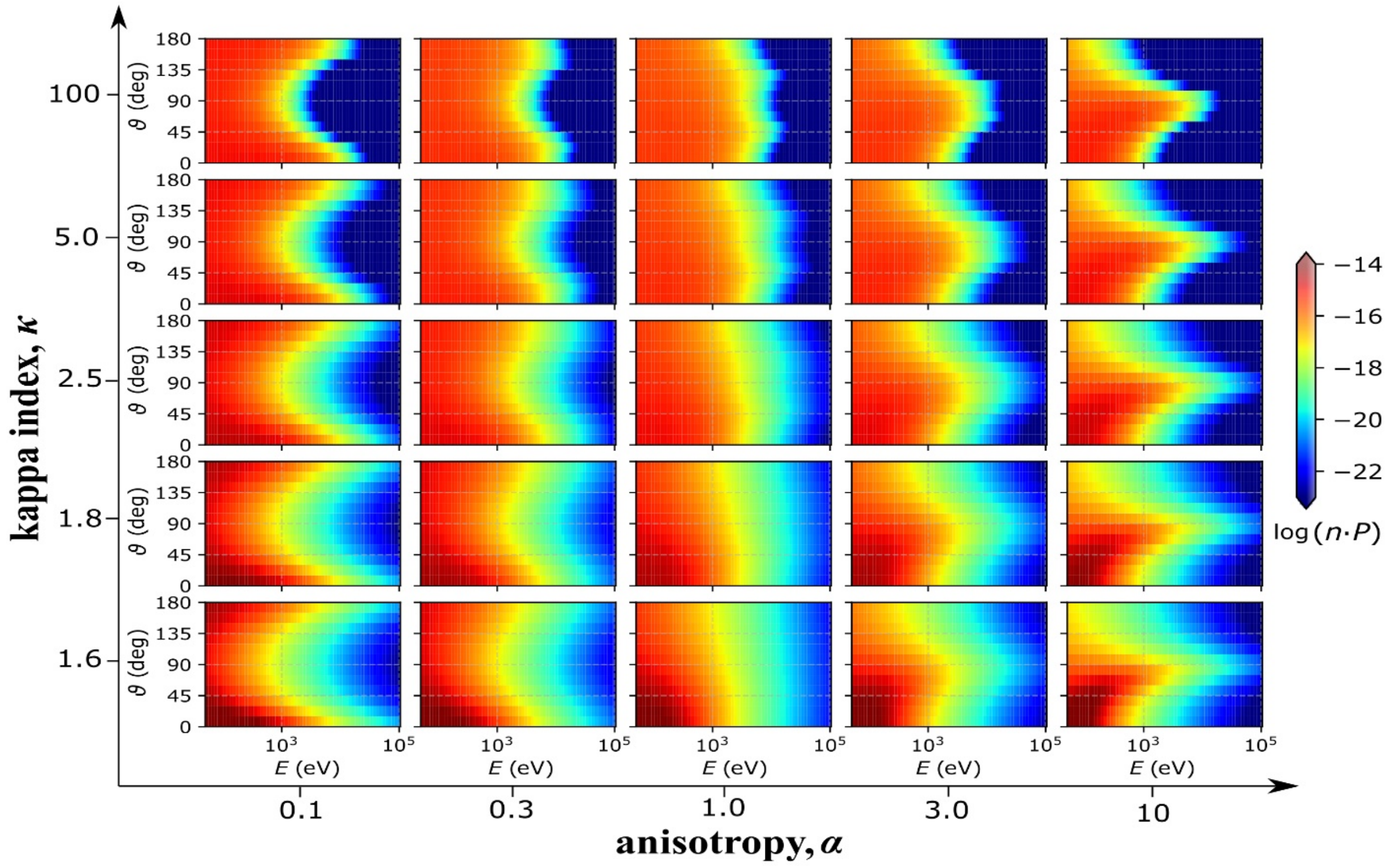

**Figure 14.** Azimuth-independent anisotropic kappa distribution of energy $E$ and pitch angle $\vartheta$, in an arbitrary reference frame, in the case of heterogeneous correlations with $\kappa^{\text{int}}$=1.5, Eq.(77a), plotted for $\vartheta_{\text{b}}=0^{0}$, $u_{\text{b}}$=5000 km/s ($E_{\text{b}}$ ~70 eV), $T$=$10^{7}$ K (~0.86 keV), and various anisotropies $\alpha$ and kappa indices $\kappa$.

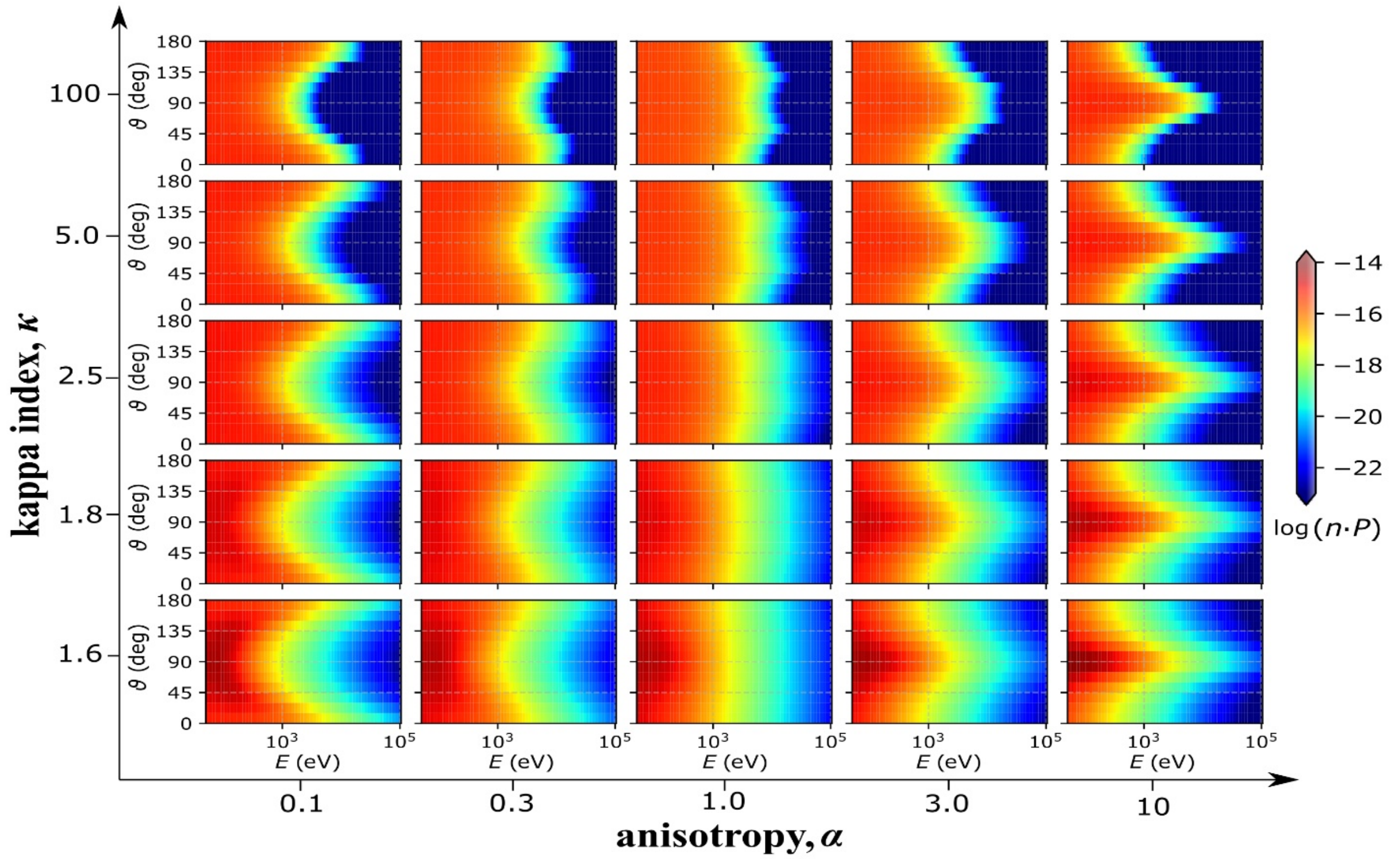

**Figure 15.** Similar to Figure 14 but for $\vartheta_{\text{b}}=90^{0}$.

## 7. Conclusions

In this paper we developed the theoretical basis for the connection of correlations among particles' velocity components with the variety of types of anisotropic kappa distributions. The correlation coefficient between two particle kinetic energies is one-to-one connected to the kappa index, the parameter that labels and governs the kappa distributions. The kappa index and the temperature constitute two intensive parameters characterizing the thermodynamics of the system. Similar to the kinetically definition of temperature, stated by the mean particle kinetic energy per half the degrees of freedom (dof), the kinetically definition of kappa is stated by the correlation (that is, the normalized covariance) of the kinetic energy per half dof.

We derived the correlation coefficient of anisotropic kappa distributions and showed how this is connected to the effective dimensionality. In particular, we were able to determine the effective dimensionality $d_{\mathrm{eff}}$ and express it as a function of the anisotropy $\alpha$, satisfying the expected limiting behavior of $d_{\mathrm{eff}}=2$ for $\alpha=0$, $d_{\mathrm{eff}}=3$ for $\alpha=1$, and $d_{\mathrm{eff}}=1$ for $\alpha\to\infty$. Then, it was straightforward to connect the adiabatic polytropic index $\gamma$ to the anisotropy $\alpha$.

Having established the importance of correlation among particles in the formulation of anisotropic kappa distributions, we proceeded and show how the variety of anisotropic formulations can be systematically defined and determined through the concept of correlations. The generalization of the anisotropic kappa distributions was developed by considering the appropriate homogeneous/heterogeneous correlations among the particles' velocity components. The derived distributions mediate the main two types of anisotropic kappa distributions, where the first considers equal correlations among particles velocity components, while the second considers zero correlation among different velocity components.

The generalization of the anisotropic kappa distributions is achieved within the framework of nonextensive statistical mechanics and its connection to the statistics of kappa distributions. There is a certain functional form that characterizes the partition of 2D joint kappa distribution into its two marginal 1D kappa distributions. This relationship leads to a statistical correlation, generalizing the relationship that leads to the statistical independence of Maxwell-Boltzmann distributions. We developed and examined the anisotropic distributions in the comoving reference frame, but also in an arbitrary S/C frame, expressed in terms of the energy, pitch angle, and azimuth, including also the case where the azimuth is not measured, thus the distribution is averaged over the azimuth. The distributions have a single maximum in the comoving frame, but more complicated graphs several peaks may characterize the same distributions in the arbitrary frame expressed in terms of energy and pitch angle.

In summary, the paper:

(i) determined the correlation coefficient and the involved effective dimensionality of anisotropic distributions with homogeneous/heterogeneous correlations;

(ii) formulated the connection of the adiabatic polytropic index with the anisotropy;

(iii) characterized and studied the homogeneous/heterogeneous correlations among the particles velocity components;

(iv) formulated the correlation relationship that characterizes the partition of 2D joint kappa distribution into the two marginal 1D kappa distributions, as emerges from nonextensive statistical mechanics;

(v) generalized the formulae of anisotropic distributions, based on the types of homogeneous/heterogeneous correlations;

(vi) described and examined the anisotropic kappa distributions in (i) the comoving reference frame with respect to the velocity components, (ii) arbitrary S/C frame with respecto to the triplet of energy, pitch angle, and azimuth, and (iii) the more complicate form of azimuth independent distributions with respect to energy and pitch angle.

Having developed the possible anisotropic kappa distributions for any reference frame, it is straightforward to apply this well-grounded "toolbox" for future reference in data analyses of space plasma populations such as electrons with Jovian Auroral Distributions Experiment (JADE) onboard the Juno Mission at Jupiter (McComas et al. 2017; Allegrini et al. 2017, 2020a; 2020b). For instance, JADE-E measurements provide the electron distribution with respect to energy and pitch angle, ignoring the azimuth values, thus the appropriate theoretical distributions to describe the collected datasets are the azimuth independent distributions with respect to energy and pitch angle, shown in §6.2.

**Acknowledgments.** The authors were partially supported by NASA's Juno mission.

**ORCID iDs**

G. Livadiotis https://orcid.org/0000-0002-7655-6019
G. Nicolaou https://orcid.org/0000-0003-3623-4928
F. Allegrini https://orcid.org/0000-0003-0696-4380

**Appendix**

Here we derive the normalization constant C of the distribution

$$\mathrm{C}\cdot\theta_\perp^{-2}\theta_\parallel^{-1}\cdot\int_0^\infty\int_{-\infty}^{+\infty}\left\{-1+\left(1+\frac{1}{\kappa-\frac{3}{2}}\cdot\frac{u_\perp^2}{\theta_\perp^2}\right)^{\frac{\kappa+\frac{1}{2}}{\kappa^{\mathrm{int}}+\frac{1}{2}}}+\left(1+\frac{1}{\kappa-\frac{3}{2}}\cdot\frac{u_\parallel^2}{\theta_\parallel^2}\right)^{\frac{\kappa}{\kappa^{\mathrm{int}}}}\right\}^{-\kappa^{\mathrm{int}}-1}du_\parallel\,2\pi u_\perp du_\perp = 1\ , \qquad \text{(A1)}$$

from the normalization

$$\int_0^\infty\int_{-\infty}^{+\infty}P(u_\parallel,u_\perp;\theta_\parallel,\theta_\perp,\kappa)\,du_\parallel\,2\pi u_\perp du_\perp = 1\ . \qquad \text{(A2)}$$

Setting $x_\perp^2\equiv u_\perp^2/[(\kappa-\frac{3}{2})\theta_\perp^2]$, $x_\parallel^2\equiv u_\parallel^2/[(\kappa-\frac{3}{2})\theta_\parallel^2]$, we find

$$\mathrm{C}^{-1}=4\pi(\kappa-\tfrac{3}{2})^{\frac{3}{2}}\cdot\int_0^\infty\int_0^\infty\left[-1+(1+x_\perp^2)^{\frac{\kappa+\frac{1}{2}}{\kappa^{\mathrm{int}}+\frac{1}{2}}}+(1+x_\parallel^2)^{\frac{\kappa}{\kappa^{\mathrm{int}}}}\right]^{-\kappa^{\mathrm{int}}-1}dx_\parallel x_\perp dx_\perp\ . \qquad \text{(A3)}$$

Next, we set $y_\perp\equiv(1+x_\perp^2)^{\frac{\kappa+\frac{1}{2}}{\kappa^{\mathrm{int}}+\frac{1}{2}}}-1$ and $y_\parallel\equiv(1+x_\parallel^2)^{\frac{\kappa}{\kappa^{\mathrm{int}}}}-1$; thus, we have $x_\perp dx_\perp=\frac{1}{2}\frac{\kappa^{\mathrm{int}}+\frac{1}{2}}{\kappa+\frac{1}{2}}\cdot(1+y_\perp)^{\frac{\kappa^{\mathrm{int}}+\frac{1}{2}}{\kappa+\frac{1}{2}}-1}\cdot dy_\perp$, $dx_\parallel=\frac{1}{2}\frac{\kappa^{\mathrm{int}}}{\kappa}\cdot(1+y_\parallel)^{\frac{\kappa^{\mathrm{int}}}{\kappa}-1}\cdot\left[(1+y_\parallel)^{\frac{\kappa^{\mathrm{int}}}{\kappa}}-1\right]^{-\frac{1}{2}}\cdot dy_\parallel$, and we find

$$\mathrm{A}^{-1}=\pi(\kappa-\tfrac{3}{2})^{\frac{3}{2}}\cdot\frac{\kappa^{\mathrm{int}}}{\kappa}\cdot\frac{\kappa^{\mathrm{int}}+\frac{1}{2}}{\kappa+\frac{1}{2}}$$
$$\times\int_0^\infty\left\{\int_0^\infty(1+y_\perp+y_\parallel)^{-\kappa^{\mathrm{int}}-1}(1+y_\perp)^{\frac{\kappa^{\mathrm{int}}+\frac{1}{2}}{\kappa+\frac{1}{2}}-1}dy_\perp\right\}(1+y_\parallel)^{\frac{\kappa^{\mathrm{int}}}{\kappa}-1}\left[(1+y_\parallel)^{\frac{\kappa^{\mathrm{int}}}{\kappa}}-1\right]^{-\frac{1}{2}}dy_\parallel\ . \qquad \text{(A4)}$$

The integration over $y_\perp$ leads to the Gauss hypergeometric function (Gradshteyn & Ryzhik 1965, Ch. 3.197-5, p.286), i.e.,

$$\int_0^\infty(1+y_\perp+y_\parallel)^{-\kappa^{\mathrm{int}}-1}(1+y_\perp)^{\frac{\kappa^{\mathrm{int}}+\frac{1}{2}}{\kappa+\frac{1}{2}}-1}dy_\perp =$$
$$\frac{1}{\kappa^{\mathrm{int}}+1-\frac{\kappa^{\mathrm{int}}+\frac{1}{2}}{\kappa+\frac{1}{2}}}\cdot(1+y_\parallel)^{-\kappa^{\mathrm{int}}-1}\cdot{}_2F_1\left(\kappa^{\mathrm{int}}+1,1;\kappa^{\mathrm{int}}+2-\frac{\kappa^{\mathrm{int}}+\frac{1}{2}}{\kappa+\frac{1}{2}};\frac{y_\parallel}{1+y_\parallel}\right). \qquad \text{(A5)}$$

Hence,

$$\mathrm{C}^{-1}=\pi(\kappa-\tfrac{3}{2})^{\frac{3}{2}}\cdot\frac{\kappa^{\mathrm{int}}}{\kappa}\cdot\frac{\kappa^{\mathrm{int}}+\frac{1}{2}}{\kappa^{\mathrm{int}}(\kappa-\frac{1}{2})+\kappa}$$
$$\times\int_0^\infty(1+y_\parallel)^{\frac{\kappa^{\mathrm{int}}}{\kappa}-\kappa^{\mathrm{int}}-2}\left[(1+y_\parallel)^{\frac{\kappa^{\mathrm{int}}}{\kappa}}-1\right]^{-\frac{1}{2}}\cdot{}_2F_1\left(\kappa^{\mathrm{int}}+1,1;\kappa^{\mathrm{int}}+2-\frac{\kappa^{\mathrm{int}}+\frac{1}{2}}{\kappa+\frac{1}{2}};\frac{y_\parallel}{1+y_\parallel}\right)dy_\parallel\ , \qquad \text{(A6)}$$

or

$$\mathrm{C}^{-1}=\pi(\kappa-\tfrac{3}{2})^{\frac{3}{2}}\cdot\frac{\kappa^{\mathrm{int}}}{\kappa}\cdot\frac{\kappa^{\mathrm{int}}+\frac{1}{2}}{\kappa^{\mathrm{int}}(\kappa-\frac{1}{2})+\kappa}\cdot I(\kappa,\kappa^{\mathrm{int}})\ . \qquad \text{(A7a)}$$

with

$$I(\kappa,\kappa^{\mathrm{int}})\equiv\int_0^\infty(1+y_\parallel)^{\frac{\kappa^{\mathrm{int}}}{\kappa}-\kappa^{\mathrm{int}}-2}\left[(1+y_\parallel)^{\frac{\kappa^{\mathrm{int}}}{\kappa}}-1\right]^{-\frac{1}{2}}{}_2F_1\left(\kappa^{\mathrm{int}}+1,1;\kappa^{\mathrm{int}}+2-\frac{\kappa^{\mathrm{int}}+\frac{1}{2}}{\kappa+\frac{1}{2}};\frac{y_\parallel}{1+y_\parallel}\right)dy_\parallel\ . \qquad \text{(A7b)}$$

Then, setting $u\equiv(1+y_\parallel)^{-1}$, Eq.(A7b) becomes

$$I(\kappa,\kappa^{\rm int}) = \int_0^1 u^{\kappa^{\rm int}(1-\frac{1}{2\kappa})}\left(1-u^{\frac{\kappa^{\rm int}}{\kappa}}\right)^{-\frac{1}{2}} {}_2F_1\left(\kappa^{\rm int}+1,1;\kappa^{\rm int}+2-\tfrac{\kappa^{\rm int}+\frac{1}{2}}{\kappa+\frac{1}{2}};1-u\right)du \quad . \tag{A8}$$

Also, we expand $(1-x)^{-\frac{1}{2}}$ in terms of $x = u^{\kappa^{\rm int}/\kappa}$,

$$I(\kappa,\kappa^{\rm int}) = \sum_{m=0}^{\infty}\tfrac{(2m)!}{4^m(m!)^2}\cdot a_m(\kappa,\kappa^{\rm int}) \text{ , with} \tag{A9a}$$

$$a_m(\kappa,\kappa^{\rm int}) \equiv \int_0^1 u^{\kappa^{\rm int}+\frac{\kappa^{\rm int}}{\kappa}(m-\frac{1}{2})} {}_2F_1\left(\kappa^{\rm int}+1,1;\kappa^{\rm int}+2-\tfrac{\kappa^{\rm int}+\frac{1}{2}}{\kappa+\frac{1}{2}};1-u\right)du \quad , \tag{A9b}$$

because

$$(1-x)^{-\frac{1}{2}} = \sum_{m=0}^{\infty}\binom{-\frac{1}{2}}{m}(-1)^m x^m \text{ , for } 0<x<1\text{, and ,} \tag{A10a}$$

$$\binom{-\frac{1}{2}}{m} = \frac{(-1)^m(2m)!}{4^m(m!)^2} \quad . \tag{A10b}$$

Hence, we find

$$\begin{aligned} a_m(\kappa,\kappa^{\rm int}) &= \int_0^1 (1-u)^{[\kappa^{\rm int}+1+\frac{\kappa^{\rm int}}{\kappa}(m-\frac{1}{2})]-1} {}_2F_1\left(\kappa^{\rm int}+1,1;\kappa^{\rm int}+2-\tfrac{\kappa^{\rm int}+\frac{1}{2}}{\kappa+\frac{1}{2}};u\right)du \\ &= [\kappa^{\rm int}+1+\tfrac{\kappa^{\rm int}}{\kappa}(m-\tfrac{1}{2})]^{-1}\cdot {}_3F_2\left(\kappa^{\rm int}+1,1,1;\kappa^{\rm int}+2-\tfrac{\kappa^{\rm int}+\frac{1}{2}}{\kappa+\frac{1}{2}},\kappa^{\rm int}+2+\tfrac{\kappa^{\rm int}}{\kappa}(m-\tfrac{1}{2});1\right) , \end{aligned} \tag{A11}$$

(see: Gradshteyn & Ryzhik 1965, Ch. 7.512-5, p.849), or

$$I(\kappa,\kappa^{\rm int}) = \sum_{m=0}^{\infty}\frac{(2m)!}{4^m(m!)^2[\kappa^{\rm int}+1+\frac{\kappa^{\rm int}}{\kappa}(m-\frac{1}{2})]}\cdot {}_3F_2\left(\kappa^{\rm int}+1,1,1;\kappa^{\rm int}+2-\tfrac{\kappa^{\rm int}+\frac{1}{2}}{\kappa+\frac{1}{2}},\kappa^{\rm int}+2+\tfrac{\kappa^{\rm int}}{\kappa}(m-\tfrac{1}{2});1\right) . \tag{A12}$$

Finally, we arrive at

$$\begin{aligned} \mathrm{C} &= \pi^{-1}\cdot\frac{\kappa}{\kappa^{\rm int}}\cdot\frac{\kappa^{\rm int}(\kappa-\frac{1}{2})+\kappa}{(\kappa-\frac{3}{2})^{\frac{3}{2}}(\kappa^{\rm int}+\frac{1}{2})} \\ &\times\left[\sum_{m=0}^{\infty}\frac{(2m)!}{4^m(m!)^2[\kappa^{\rm int}+1+\frac{\kappa^{\rm int}}{\kappa}(m-\frac{1}{2})]}\cdot {}_3F_2\left(\kappa^{\rm int}+1,1,1;\kappa^{\rm int}+2-\tfrac{\kappa^{\rm int}+\frac{1}{2}}{\kappa+\frac{1}{2}},\kappa^{\rm int}+2+\tfrac{\kappa^{\rm int}}{\kappa}(m-\tfrac{1}{2});1\right)\right]^{-1} , \end{aligned} \tag{A13a}$$

or

$$\begin{aligned} \mathrm{C} &= \pi^{-1}\cdot\frac{\kappa[\kappa^{\rm int}(\kappa-\frac{1}{2})+\kappa]}{(\kappa-\frac{3}{2})^{\frac{3}{2}}(\kappa^{\rm int}+\frac{1}{2})}\cdot\frac{(\kappa^{\rm int}-1)!}{[\kappa^{\rm int}+1-\frac{\kappa^{\rm int}+\frac{1}{2}}{\kappa++\frac{1}{2}}]!} \\ &\times\left[\sum_{m=0}^{\infty}\frac{(2m)![\kappa^{\rm int}+\frac{\kappa^{\rm int}}{\kappa}(m-\frac{1}{2})]!}{4^m(m!)^2}\cdot\sum_{n=0}^{\infty}\frac{(\kappa^{\rm int}+n)!n!}{[\kappa^{\rm int}+1-\frac{\kappa^{\rm int}+\frac{1}{2}}{\kappa++\frac{1}{2}}+n]![\kappa^{\rm int}+1+\frac{\kappa^{\rm int}}{\kappa}(m-\frac{1}{2})+n]!}\right]^{-1} . \end{aligned} \tag{A13b}$$